\documentclass[a4paper,11pt]{article}
\pdfoutput=1 

\usepackage{jheppub} 
\usepackage[T1]{fontenc} 

\usepackage[latin1]{inputenc}
\usepackage[english]{babel}

\usepackage{amssymb,amsthm,cancel,hyperref,graphicx,xcolor}
\usepackage{picinpar,graphicx,xypic}
\usepackage{booktabs}
\usepackage{mathrsfs}
\usepackage{amsfonts}
\usepackage{latexsym}
\usepackage{booktabs,graphicx,hyperref,epsfig,multirow}
\usepackage{bm}
\allowdisplaybreaks
\def\bea{\begin{align}}
\def\eea{\end{align}}
\def\beq{\begin{equation}}
\def\eeq{\end{equation}}
\def\ba{\begin{eqnarray}}
\def\ea{\end{eqnarray}}
\def\be{\begin{equation}}
\def\ee{\end{equation}}
\definecolor{darkgreen}{HTML}{008000}
\newcommand{\sss}{\scriptscriptstyle\rm}
\newcommand{\muf}{\mu_{\rm\sss F}}
\newcommand{\mur}{\mu_{\rm\sss R}}

\renewcommand{\Re}{\mathrm{Re}}

\newcommand{\plus}[1]{\left[ #1 \right]_+}

\newcommand{\gammae}{\gamma_{\scriptscriptstyle E}}
\newcommand{\Ord}{\mathcal{O}}
\newcommand{\gsim}{\gtrsim}

\newcommand{\as}{\alpha_s}

\def\({\left(}
\def\){\right)}
\def\[{\left[}
\def\]{\right]}

\def    \hepph  #1 {{\tt hep-ph/#1}}
\def    \hepex  #1 {{\tt hep-ex/#1}}
\long\def\symbolfootnote[#1]#2{\begingroup%
\def\thefootnote{\fnsymbol{footnote}}\footnote[#1]{#2}\endgroup}

\numberwithin{equation}{section}
\graphicspath{{TopQuarkImage/}}

\renewcommand{\(}{\left(}
\renewcommand{\)}{\right)}
\renewcommand{\Re}{\mathrm{Re}\:}
\newcommand{\MSbar}{$\overline{\rm MS}$}

\newcommand{\sigg}{\hat{\sigma}}
\newcommand{\plusq}[1]{\left[#1\right]_+}
\newcommand{\Ca}{C_{\rm\sss A}}
\newcommand{\Cf}{C_{\rm\sss F}}
\newcommand{\Nf}{n_f}

\newcommand{\gammapres}{\gamma_{\rm res}^+}
\newcommand{\dotgammapres}{\dot\gamma_{\rm res}^+}

\newcommand{\kt}{k_{\rm\sss T}}

\newcommand{\sqmatr}[9]{\left(
    \begin{array}[c]{ccc}
      #1 & #2 & #3 \\ 
      #4 & #5 & #6 \\ 
      #7 & #8 & #9 
    \end{array}
  \right)}
\graphicspath{{TopQuarkImage/}}
\renewcommand{\(}{\left(}
\renewcommand{\)}{\right)}
\renewcommand{\Re}{\mathrm{Re}\:}

\title{\boldmath Top Quark Pair Production beyond NNLO}

\author[a]{Claudio Muselli,}
\author[b]{Marco Bonvini,}
\author[a]{Stefano Forte,}
\author[c]{Simone Marzani}
\author[d]{and Giovanni Ridolfi}

\affiliation[a]{TIF Lab, Dipartimento di Fisica, Universit\`a di Milano,  and
INFN, Sezione di Milano,\\ Via Celoria 16, I-20133 Milano, Italy}
\affiliation[b]{Rudolf Peierls Centre for Theoretical Physics,
  University of Oxford,\\ 1 Keble Road, OX1 3NP, Oxford, UK}
\affiliation[c]{Center for Theoretical Physics, Massachusetts Institute
of Technology,\\ 77 Massachusetts Avenue, Cambridge, MA 02139, USA}
\affiliation[d]{Dipartimento di Fisica, Universit\`a di Genova and 
INFN, Sezione di Genova,\\
Via Dodecaneso 33, I-16146 Genova, Italy}

\emailAdd{claudio.muselli@mi.infn.it}
\emailAdd{marco.bonvini@physics.ox.ac.uk}
\emailAdd{stefano.forte@mi.infn.it}
\emailAdd{smarzani@mit.edu}
\emailAdd{giovanni.ridolfi@ge.infn.it}

\preprint{
\begin{flushright}
TIF-UNIMI-2015-4\\
OUTP-15-09P\\
MIT-CTP 4656
\end{flushright}
}

\abstract{
We construct an approximate expression for the total cross section for
the production of a heavy quark-antiquark pair in hadronic collisions
at next-to-next-to-next-to-leading order (N$^3$LO) in $\as$. We
use a technique which exploits the analyticity of the
Mellin space cross section, and the information on its singularity
structure coming from large $N$  (soft gluon, Sudakov)
and small $N$ (high energy, BFKL) all order resummations, previously
introduced  and used in the case of Higgs production.
We validate our method by comparing to available exact results up to
NNLO. We find that N$^3$LO corrections increase the predicted 
top pair cross section at the LHC by about 4\% over the NNLO.}

\begin{document} 
\maketitle
\flushbottom

\section{Introduction}
\label{sec:intro}

The lack of discovery of any new physics signal so far has made of the
Large Hadron Collider (LHC) even more a precision machine than it ever
was. Because the LHC is a hadron collider, the largest uncertainties
are related to nucleon structure, i.e.\ parton distributions, and
higher order QCD corrections. This has consequently led to
enormous progress in the computation of higher order corrections to
QCD processes in the last several years, recently leading to the
publication of first  N$^3$LO results
for a hadron collider process~\cite{Anastasiou:2015ema}.

In Ref.~\cite{Ball:2013bra} some of us have
proposed a general methodology for the determination of approximate
expressions for higher-order corrections to QCD processes. The basic
idea is that QCD corrections to many important hard processes are
known to all orders in the strong coupling in two opposite limits: the
high energy limit, in which the available center-of-mass energy is
much greater than the invariant mass of the final state, and the
soft  limit, in which the invariant mass is close to
threshold. By performing a Mellin  transform, this knowledge can be
turned into information on the singularities of the hard (partonic)
Mellin-space
cross section to any given finite order, viewed as an analytic
function of the Mellin complex variable: the soft limit
determines the singularity at infinity, and the high-energy limit
determines the rightmost pole on the real axis. It is then possible to
reconstruct an approximate form of the function by exploiting this
knowledge: indeed, knowledge of all the singularities would determine the
function completely.

The case of Higgs production, to which the methodology of
Ref.~\cite{Ball:2013bra} was first
applied~\cite{Ball:2013bra,Bonvini:2014jma,Bonvini:2014joa}, is
particularly simple in many respects: the leading-order process has
fixed kinematics (only one scalar colorless particle in the final
state), and the total cross section is known~\cite{Kramer:1996iq} to
be dominated by its threshold limit. The validation of the methodology
which comes from checking that it provides a good approximation to
known results in the case of Higgs is thus not necessarily the most
compelling.

In this paper we turn our attention to heavy quark production: a
process which is rather more complicated in terms of kinematics and color
structure, with the goal of performing a more stringent test of our
methodology. Next-to-leading (NLO) QCD corrections to this process
were computed long ago~\cite{Nason:1987xz,Mangano:1991jk}, though
fully analytic expressions only became available much more
recently~\cite{Czakon:2008ii}, and were a first necessary step towards
the full determination of the NNLO result which was very recently
achieved~\cite{Baernreuther:2012ny,Czakon:2012zr,Czakon:2012pz,Czakon:2013goa}.
These results revealed that the physical-space partonic cross section
has a particularly complex singularity structure, which raises the
question of how this relates to the Mellin-space singularities which
are used for the approximation of Ref.~\cite{Ball:2013bra}, and also,
that existing attempts at an approximate NNLO
determination~\cite{Moch:2012mk} were rather off the mark.

Here, we will present an approximate determination of the N$^3$LO QCD
corrections to the total cross section for heavy quark production.
In  Section~\ref{sec:Singularities} we will review the singularity
structure of the heavy quark production partonic cross section both in
physical and in Mellin  space.  Specifically, we will discuss 
the relationship between
the total cross section and the invariant mass distribution, to which
the methods of  Ref.~\cite{Ball:2013bra} apply,
and show that the peculiar singularity structure of the physical-space
cross section leads nevertheless to the standard Mellin-space
singularities. The rest of our treatment closely follows that of Higgs
production: in Sects.~\ref{sec:largeN} and \ref{sec:smallN}
respectively we present the derivation of the soft and high-energy
singularities from the respective resummations. They are combined in
Section~\ref{sec:approx} where we present our final result: first, we
compare the approximation obtained through 
our procedure to the known exact result up to NNLO, and then we
present our final approximate N$^3$LO both at the parton and the
hadron level.

\section{Analytic structure of the partonic cross section} 
\label{sec:Singularities}

The analytic structure of the cross section for the production of a
heavy quark pair potentially differs from that for Higgs production
discussed in Ref.~\cite{Ball:2013bra} for two different reasons,
thereby potentially hampering, or requiring some adaptation of the
methods used in that reference.
First, the known~\cite{Czakon:2008ii} fact that the partonic cross
section for heavy quark production already at NLO
has unphysical singularities in the
complex plane of its kinematic variables may suggest that the
singularity structure of its Mellin transform also does not follow the
general pattern discussed in  Ref.~\cite{Ball:2013bra}. Second, as we
shall see in more detail below, the analytic structure discussed in 
Ref.~\cite{Ball:2013bra} is generic for partonic differential
cross sections, while the case we are presently considering is that of
a total cross section. We will see that the first
issue is of no concern, while the second requires a careful discussion
of the relation of the total heavy quark production
cross section  to the corresponding invariant mass distribution.

We will address the two issues in turn, after a quick review of
standard notation and general results.

\subsection{Notation and general results}
\label{sect:general}

The total cross section for the production of a heavy 
quark-antiquark pair can be written in factorized form as
\beq
\label{eq:hqpcross}
\sigma(m^2,\rho_h)=\rho_h\sum_{ij}\int_{\rho_h}^1\frac{d\rho}{\rho}\,
\mathscr{L}_{ij}\(\frac{\rho_h}{\rho},\muf^2\)\frac{1}{\rho}
\hat{\sigma}_{ij}\(m^2,\rho,\as(\mur^2),\frac{m^2}{\muf^2},\frac{m^2}{\mur^2}\)
\eeq
where $\rho_h=4m^2/s$, $s$ is the hadronic center of mass
energy, $\mathscr{L}_{ij}(\rho,\muf^2)$ are parton luminosities, $m$ is
the mass of the heavy quark, $\muf,\mur$ are factorization and
renormalization scales, and the scaling variable is $\rho=4m^2/\hat s$, with
$\hat s$ the partonic center of mass energy.  To simplify notations, in the
following we will not indicate explicitly the dependence of parton
luminosities on the factorization scale $\muf$ and that
of the partonic cross section on the strong coupling and on
$\muf,\mur$. 

The partonic cross sections $\hat\sigma_{ij}$ 
admit a  perturbative expansion in powers of the QCD coupling:
\beq\label{eq:sigmahatseries}
\hat{\sigma}_{ij}(m^2,\rho)
=\frac{\as^2}{m^2}\left[\hat{\sigma}^{\(0\)}_{ij}(\rho)
+\as \hat{\sigma}^{(1)}_{ij}(\rho)
+\as^2\hat{\sigma}^{(2)}_{ij}(\rho)
+\as^3\hat{\sigma}^{(3)}_{ij}(\rho)
+\Ord(\as^4)\right],
\eeq
where, thanks to the  $m^{-2}$ prefactor,  the coefficients
$\hat{\sigma}^{(k)}_{ij}(\rho)$ are dimensionless.
Eq.~\eqref{eq:hqpcross} is in the form of a convolution product, so
its Mellin transform
\beq
\sigma(m^2,N)=\int_0^1d\rho_h\, \rho_h^{N-2}\sigma(m^2,\rho_h)
\eeq
factorizes in terms of the Mellin space luminosity and partonic cross 
section function, defined respectively as
\begin{align}
\label{eq:defMellinlum}
\mathscr{L}(N)&=\int_0^1 d\rho\, \rho^{N-1}\mathscr{L}(\rho)\\
\label{eq:defMellincross}
\hat{\sigma}_{ij}(m^2,N)&=\int_0^1 d\rho\, \rho^{N-2}
\hat{\sigma}_{ij}(m^2,\rho),
\end{align}
according to
\beq\label{eq:melfac}
\sigma(m^2,N)=\sum_{ij}\mathscr{L}_{ij}(N)\hat{\sigma}_{ij}(m^2,N).
\eeq
If the Mellin transform integral has a finite convergence abscissa,
the $N$-space partonic cross section is an analytic function of the
complex variable $N$, defined by the integral representation
Eq.~\eqref{eq:defMellincross} to the right of the convergence
abscissa, and by analytic continuation elsewhere. Therefore, it is
fully determined by the knowledge of its singularities.

In this paper, we will concentrate on the $gg$ partonic channel, which
is the most relevant (in the $\overline{\rm MS}$ scheme) at LHC
energies, while we leave the study of other parton subprocesses for
future work.  For this reason in the following we will drop the parton
indices $ij$, with the understanding that $i=j=g$ in all parton
luminosities and partonic cross sections.

The singularity structure of a generic differential partonic cross section is
relatively simple. The singularity at infinity 
is determined by soft gluon radiation, which leads to a growth of the
cross section with increasingly high powers of $\ln N$ at higher
perturbative orders. For finite $N$, 
 singularities away from the real axis are not allowed, as they would
have to come in pairs and would thus lead to oscillatory behaviour of
the cross section at high energy. From Regge theory one expects that the
rightmost singularity on the real axis  is a multiple pole located at
$N=1$~\cite{Lipatov:polo}, with further multiple poles along the real
axis at $N=0,-1,-2,\dots$. This expectation is confirmed by explicit
fixed-order calculations. 

While knowledge of the residues of all poles is required in order to
fully determine the function $\hat{\sigma}(m^2,N)$, its behaviour 
in the region $1< \Re N <\infty$ is mostly 
controlled by the rightmost pole at $N=1$, with poles further to the
left having an increasingly small impact. Using the
saddle point approximation one can show that the
hadronic cross section is mostly determined by the behaviour of
the partonic cross section
$\hat{\sigma}(m^2,N)$ in the vicinity of a single (saddle) value of
$N$ on the real axis, to the right of the rightmost
singularity~\cite{Bonvini:2012an, Ball:2013bra} (see
also Sect.~\ref{sect:saddle}). 
In Ref.~\cite{Ball:2013bra} it was suggested that knowledge of the
rightmost singularity on the real axis and of the singularity at
infinity is sufficient to determine the partonic cross section in the
region where the saddle-point is located with reasonable accuracy.

As mentioned in the introduction, however, in the specific case of
heavy quark production, we encounter two difficulties. 
The first, is related to the fact that the physical momentum-space coefficient
function has unphysical singularities. The second is due to the fact
that the partonic cross sections $\hat{\sigma}_{ij}(m^2,N)$
Eq.~(\ref{eq:melfac}) vanish as $N\to \infty$, rather than growing
logarithmically.

We will address both issues in turn. The first issue turns out to be
of no concern: the analytic structure in Mellin space is unaffected by
the unphysical momentum-space singularities. The second issue instead
is due to the fact that the reason why the partonic cross sections
discussed in Ref.~\cite{Ball:2013bra} grow as $N\to\infty$ is that
they are distributions, rather than ordinary functions: indeed,
elementary properties of Laplace transforms imply that the Mellin
transform of an ordinary function, if it exists, must vanish as
$N\to\infty$. Now, the integral of a distribution is an ordinary
function, so it is clear that the behaviour of
Ref.~\cite{Ball:2013bra} can only hold for partonic cross sections
which are sufficiently differential. As we shall see below, it is the
invariant-mass distribution which behaves at large $N$ in the way
discussed in Ref.~\cite{Ball:2013bra}. However, by exploiting the
relation between total cross section and invariant-mass distribution, 
 it is possible to relate their respective large-$N$ behaviors, and
 define a coefficient
function whose singularity structure follows the general pattern
discussed above.

\subsection{Unphysical singularities in momentum space}
\label{sect:sings}

The
perturbative coefficients $\hat\sigma_{ij}^{(k)}(\rho)$ display a
class of spurious singularities in the complex $\rho$ plane outside
the physical region $0<\rho\leq 1$. However, 
after Mellin transformation, even in the presence of
such spurious singularities in $\rho$ space, the result has the
expected analytic structure, as we now show.

As an example, we present the case of NLO corrections of top pair
production in the $gg$ channel. In Ref.~\cite{Czakon:2008ii}, the
complete analytic form of $\hat{\sigma}^{\(1\)}_{gg}\(\rho\)$ is
computed, and it contains, in addition to the usual (physical)
singularities in $\rho=1$ and $\rho=0$, four further singularities
(branch points) located at $\rho=-4$, $\rho=4$, $\rho=-1$, $\rho =
-\frac{1}{4}$ (see in particular the functions $F_i(\rho)$,
$i=1,\ldots,4$ of section $4$ and the discussion of section $5$ of
Ref.~\cite{Czakon:2008ii}). For simplicity we focus our attention only
on the first spurious singularity; similar considerations hold for all
the others.

The function
\begin{align}
\label{eq:F2}
F_2\(\rho\)&=\int_{x(\rho)}^1dz\,f_2(z)
\\
f_2(z)&=-\frac{\(2z+3\)\(12\ln z\ln \frac{1+z}{\sqrt{z}}
+3\ln^2 z +12 \mathrm{Li}_2\(-z\)+\pi^2\)}{4\(z^2+3z+1\)},
\\
x(\rho)&=\frac{1-\sqrt{1-\rho}}{1+\sqrt{1-\rho}}
\end{align}
has a logarithmic branch cut starting at $\rho=-4$, because the integrand 
has a simple pole in
\beq
z=\frac{\sqrt{5}-3}{2}=x(-4).
\eeq

However, the Mellin transform is
\beq
F_2(N)=\int_0^1d\rho\,\rho^{N-2} \int_{x(\rho)}^1dz\,f_2(z)
=\int_0^1dz\,f_2\(z\)\int_0^{\frac{4z}{\(1+z\)^2}} d\rho\, \rho^{N-2}:
\eeq
the $\rho$ integral is convergent for $\Re N > 1$, and the result
\beq
F_2(N)=\frac{4^{N-1}}{N-1}\int_0^1dz\,z^{N-1}(1+z)^{2-2N}f_2(z)
\eeq
can be analytically continued to the whole complex $N$ plane, except
the isolated points $N=1,0,-1,-2,\dots$, where the result has simple
poles. 
This can be seen for example by repeatedly integrating by parts
using $z^{N-1}$ as a differential factor.  

This is the
analytic structure one expects for the Mellin transform of a physical
cross section. We conclude that the presences of unphysical
singularities in the $\rho$ space does not affect the general
structure the partonic cross section in $N$ space. Of course, it could
be that the presence of new structures in the partonic cross section
at higher perturbative orders affects the numerical size of the
residues of singularities, for the same reasons why it makes standard
scale-variation estimates of higher order terms unreliable.

\subsection{Total cross section and invariant-mass distribution}
\label{sec:diff}

We will first show that the invariant-mass distribution for
heavy-quark production has the large $N$ singularity structure
discussed in Ref.~\cite{Ball:2013bra} and dominated by Sudakov
radiation, then we will   prove that the coefficient function $C(N)$,
defined by factoring out the leading order
contribution in the total cross section:
\beq
\label{eq:defcoeff}
\hat{\sigma} (m^2,N)=\hat{\sigma}^{\sss LO}(m^2,N) C(N), 
\eeq 
exhibits the same Sudakov enhancement at large $N$.

The
invariant-mass distribution $\Sigma(m^2,\xi,z)$ is defined as
\beq
\Sigma(m^2,\xi,z)=\hat s\frac{d\hat\sigma}{dM^2}
(m^2,\xi,z),\label{imdef}
\eeq
which is a function of two dimensionless ratios, due to the presence
of an extra energy scale $M^2$, the invariant mass of the
quark-antiquark pair. We choose to express it as a function of
$z=M^2/\hat s$, and of $\xi=4m^2/M^2=\rho/z$, and we insert the factor
of $\hat s$ in Eq.~(\ref{imdef}) for later convenience.
The total cross section is related to the invariant-mass distribution by
\beq
\label{eq:link1}
\hat\sigma(m^2,\rho)=\int_{4m^2}^{\hat s} dM^2\,\frac{d\hat\sigma}{dM^2}
=\int_\rho^1 dz\,\Sigma\(m^2,\frac{\rho}{z},z\).
\eeq

In  Mellin space, the relation between $\hat\sigma$ and $\Sigma$ can
be obtained by computing the Mellin transform of Eq.~\eqref{eq:link1}:
\begin{align}
\label{eq:link2}
\hat\sigma(m^2,N)
&=\int_0^1d\rho\,\rho^{N-2}\int_\rho^1 dz\,\Sigma\(m^2,\frac{\rho}{z},z\)
\\
&=\int_0^1dz\,\int_0^z d\rho\,\rho^{N-2}\Sigma\(m^2,\frac{\rho}{z},z\).
\end{align}
Changing integration variable we obtain
\beq
\label{eq:link4}
\hat\sigma(m^2,N)=\int_0^1dz\,z^{N-1} \int_0^1 d\xi\, \xi^{N-2}  
\Sigma(m^2,\xi,z)=\Sigma(m^2,N-1,N).
\eeq

Using this last relation, we will show that $C(N)$ implicitly defined
in Eq.~\eqref{eq:defcoeff} has the same Sudakov singularities in the
large $N$ limit  as the invariant
mass distribution $\Sigma(m^2,\xi,N)$ in the limit $\xi\to 1$. This
limit is called the \emph{absolute threshold} limit, and it
corresponds to the double limit $\xi\to 1$ and $z\to 1$ of
$\Sigma\(m^2,\xi,z\)$ in $z$ space.  Indeed, we see from
Eq.~\eqref{eq:link1} that, when $\rho \to 1$, only the region $z\to 1$
(and hence $\xi\to 1$) contributes to the integral.  
  
The behaviour of $\Sigma(m^2,\xi,z)$ for $z$ close to one is governed
by Sudakov resummation. In analogy with Drell-Yan or resonant Higgs
production, the $\Ord(\as^n)$ perturbative coefficient for
$\Sigma(m^2,\xi,z)$ is a linear combination of the distributions
\beq
\mathcal{D}_k(z)=\plus{\frac{\ln^k(1-z)}{1-z}}, \qquad \delta(1-z)
\eeq
with $0\leq k\leq 2n-1$, plus contributions which are less singular in the
threshold limit $z\to 1$. Correspondingly,
the perturbative coefficients of
\beq
\Sigma(m^2,\xi,N)=\int_0^1 dz\,z^{N-1}\Sigma(m^2,\xi,z)
\eeq
grow at large $N$ as $\ln^kN$, $0\leq k\leq 2n$. 

The resummation of these large
logarithmic contributions has been performed in recent
years, both in momentum space with Soft Collinear Effective Theory (SCET) techniques~\cite{Ahrens:2010zv,Ahrens:2011px,Yang:2014hya}
and in Mellin space~\cite{Kidonakis:2013zqa,Kidonakis:2011ca,Kidonakis:2001nj}.
In the latter case one finds~\cite{Kidonakis:2011ca}
\begin{align}
\label{eq:Stermanfact}
\Sigma^{\rm res}(m^2,\xi,N)&=
\exp\left[G\(N\)\right]{\rm Tr\,}\Bigg\{\mathbf{H}(\xi,\as(M^2))\notag \\
&\times \exp\left[\int_0^1 dz\,\frac{z^{N-1}-1}{1-z} 
\mathbf{\Gamma_S}^\dagger(\xi,\as(M^2(1-z)^2))\right] 
\mathbf{S}\(\as\(\frac{M^2}{N^2}\)\)\notag \\
&\times \exp\left[\int_0^1 dz\,\frac{z^{N-1}-1}{1-z}
\mathbf{\Gamma_S}(\xi,\as(M^2(1-z)^2))\right]\Bigg\}
\end{align}
up to corrections which vanish in the large-$N$ limit.
Here, bold symbols are used for matrices in color space. 
In the case of heavy quark pair production in
gluon-gluon fusion, these are $3\times 3$ matrices because
there are three independent colour configurations~\cite{Kidonakis:2001nj}.
The function $G(N)$ is given by
\beq
G(N)=\int_0^1dz\,\frac{z^{N-1}-1}{1-z}\left[2\int_{\muf^2}^{M^2\(1-z\)^2} 
\frac{d\mu^2}{\mu^2} A(\as(\mu^2)) + D(\as(M^2(1-z)^2))\right]
\eeq
where $A(\as)$~\cite{Catani:2003zt,Vogt:2004mw} and
$D(\as)$~\cite{Contopanagos:1996nh} have perturbative expansions in
powers of $\as$, which are known up to $\mathcal{O}\left(
\as^3\right)$ (see for
instance~\cite{Moch:2005ba,Moch:2005ky,Laenen:2005uz}).  The soft
anomalous dimension $\mathbf{\Gamma_S}(\xi,\as)$ and the soft (matrix)
function $\mathbf{S}(\as)$ originate from soft emissions in the
presence of a heavy quark pair. Finally, the matrix function
$\mathbf{H}(\xi,\as)$ represents the hard contribution to the cross
section. NNLL accuracy is achieved by expanding the cusp anomalous
dimension $A$ up to three loops, the characteristic function $D$ and
soft anomalous dimension $\mathbf{\Gamma_S}$ up to two loops, and the
soft and hard functions $\mathbf{S},\mathbf{H}$ to one loop. Moreover
to predict the term proportional to the delta function $\delta(1-z)$
at $\Ord(\as^2)$, we just need the sum of 
the two loop contributions to the hard
and soft functions, $\mathbf{H}$ and $\mathbf{S}$,
which can be determined by matching with
NNLO calculation. 

However, what we
are interested in is the absolute threshold limit $\xi\to 1$ of
Eq.~\eqref{eq:Stermanfact}. We now show that in this limit
Eq.~\eqref{eq:Stermanfact} greatly simplifies.
For $\xi\sim 1$ we can replace $M^2$ by $4m^2$ 
in the argument of $\as$, the difference being suppressed by powers
of $1-\xi$.
Furthermore, it was noted in
Ref.~\cite{Czakon:2009zw} that in this limit the matrix
$\mathbf{\Gamma_S}(\xi,\as)$ is diagonal 
in the singlet-octet basis defined e.g.\ in Ref.~\cite{Kidonakis:2001nj}.
To order $\as^2$ it takes the form~\cite{Czakon:2009zw}
\beq
\label{expabdsthres1}
\mathbf{\Gamma_S}(\xi,\as)=-\Ca\left[
\frac{\as}{2\pi}+\(\frac{\as}{2\pi}\)^2
\(K+\zeta_3-1\)\right]\mathbf{\Pi_8}\equiv\Gamma_S(\as)\mathbf{\Pi_8},
\eeq
where 
\beq
K=\(\frac{67}{18}-\frac{\pi^2}{6}\)\Ca-\frac{10}{18}\Nf,
\eeq
and the matrix $\mathbf{\Pi_8}$ is the projector over the
octet subspace:
\beq
\mathbf{\Pi_8}=\sqmatr{0}{0}{0}{0}{1}{0}{0}{0}{1}.
\eeq
We now turn to the soft matrix $\mathbf{S}(\as)$.
By a suitable definition of $\mathbf{H}$, it can be chosen to 
be the unity matrix $\mathbf{I}$ at leading order.
Its NLO expansion, sufficient to achieve NNLL accuracy, 
is given by
\beq 
\mathbf{S}(\as)
=\mathbf{1}+\frac{\Ca}{\pi}\as\mathbf{\Pi_8}
+\Ord(\as^2).
\eeq

Thus, in the absolute threshold limit, Eq.~\eqref{eq:Stermanfact} reduces to
\begin{align}
\label{eq:Stermanfact2}
\Sigma^{\rm res}(m^2,\xi,N)&=
\exp\left[G(N)\right]{\rm Tr\,}\Bigg\{\mathbf{H}(\xi,\as(m^2))
\left[
\mathbf{1}+\frac{\Ca}{\pi}\as\(\frac{4m^2}{N^2}\)\mathbf{\Pi_8}+\Ord(\as^2)
\right]
\notag \\
&\times \exp\left[2\mathbf{\Pi_8}\int_0^1 dz\,\frac{z^{N-1}-1}{1-z}
\Gamma_S(\as(4m^2(1-z)^2))\right]\Bigg\}
\end{align}
The calculation of the
color trace in Eq.~\eqref{eq:Stermanfact2} is greatly simplified 
by noting that, for any complex number $a$,
\beq
e^{a\mathbf{\Pi_8}}=e^a\mathbf{\Pi_8}+\mathbf{\Pi_1}
\eeq
where
\beq
\mathbf{\Pi_1}+\mathbf{\Pi_8}=\mathbf{1};\qquad
\mathbf{\Pi_1}^2=\mathbf{\Pi_1};\qquad
\mathbf{\Pi_8}^2 = \mathbf{\Pi_8};\qquad
\mathbf{\Pi_1}\mathbf{\Pi_8}=\mathbf{0}.
\eeq
As a consequence, the resummed invariant mass distribution splits into the
sum of an octet and a singlet component:
\begin{align}
\label{eq:Stermanfact3}
\Sigma^{\rm res}(m^2,\xi,N)
&=\exp\left[G(N)\right]
{\rm Tr\,}\left[\mathbf{H}(\xi,\as(m^2))\mathbf{\Pi_8}\right]
\left[1+\frac{\Ca}{\pi}\as\(\frac{4m^2}{N^2}\) +\Ord(\as^2)\right]
\notag\\ &\times
\exp\left[2\int_0^1 dz\,\frac{z^{N-1}-1}{1-z}
\Gamma_S(\as(4m^2(1-z)^2))\right]
\notag\\
&+\exp\left[G(N)\right]
{\rm Tr\,}\left[\mathbf{H}(\xi,\as(m^2))\mathbf{\Pi_1}\right]
\left[1+\Ord(\as^2)\right].
\end{align}

The matrix $\mathbf{H}(\xi,\as)$ cannot be expanded
in powers of $\xi-1$, because it is proportional to
the phase-space factor $\sqrt{1-\xi}$. However,
one may factorize the leading order coefficient
as in Ref.~\cite{Czakon:2009zw}, to obtain
\beq 
\mathbf{H}(\xi,\as)=\mathbf{H}^{(0)}(\xi)
\left[1+\mathbf{H}^{(1)}(\xi)\as+\mathbf{H}^{(2)}(\xi)\as^2+\Ord(\as^3)\right].
\eeq
The coefficients $\mathbf{H}^{(i)}(\xi)$, $i\ge 1$ are now
analytic around $\xi=1$. Thus
\beq 
\label{expabdsthres3}
\mathbf{H}(\xi,\as)=\mathbf{H}^{(0)}(\xi)
\left[1+\mathbf{h}^{(1)}\as+\mathbf{h}^{(2)}\as^2+\Ord(\as^3)
+\Ord(1-\xi)
\right]
\eeq
where $\mathbf{h}^{(i)}=\mathbf{H}^{(i)}(1)$ are constant matrices.
A further simplification arises
from the particular structure of the matrix $\mathbf{H}^{(0)}(\xi,\as)$.
Indeed, one finds
\beq
\mathbf{H}^{(0)}(\xi,\as)=
\sqmatr{\hat\sigma_{\mathbf{1}}^{\sss LO}(m^2,\xi)}
{\hat\sigma_{\mathbf{1}}^{\sss LO}(m^2,\xi)}{0}
{3\hat\sigma_{\mathbf{1}}^{\sss LO}(m^2,\xi)}
{\frac{5}{2}\hat\sigma_{\mathbf{1}}^{\sss LO}(m^2,\xi)}{0}
{0}{0}
{\hat\sigma_{\mathbf{8}}^{\sss LO}(m^2,\xi)
-\frac{5}{2}{\hat\sigma_{\mathbf{1}}^{\sss LO}(m^2,\xi)}
},
\eeq
where $\hat\sigma_{\mathbf{I}}^{\sss LO}(m^2,\xi)$ are the leading-order
total cross sections in each color configuration.
Since
\beq
\lim_{\xi\to 1}
\frac{\hat\sigma_{\mathbf{8}}^{\sss LO}(m^2,\xi)}
{\hat\sigma_{\mathbf{1}}^{\sss LO}(m^2,\xi)}=\frac{5}{2};
\qquad
\hat\sigma^{\sss LO}(m^2,\xi)=
\hat\sigma_{\mathbf{1}}^{\sss LO}(m^2,\xi)
+\hat\sigma_{\mathbf{8}}^{\sss LO}(m^2,\xi)
\label{thresholdLO}
\eeq
we have
\beq
\mathbf{H}^{(0)}(\xi,\as)
=\hat\sigma^{\sss LO}(m^2,\xi)\mathbf{h}^{(0)}
+\Ord(1-\xi);
\qquad
\mathbf{h}^{(0)}=\sqmatr{\frac{2}{7}}{\frac{2}{7}}{0}
{\frac{6}{7}}{\frac{5}{7}}{0}{0}{0}{0}.
\eeq
It follows that
\beq
{\rm Tr\,}\left[\mathbf{H}(\xi,\as)\mathbf{\Pi_I}\right]
=\hat\sigma^{\sss LO}(m^2,\xi)\bar g_{\mathbf{I}}(\as)
+\Ord(1-\xi);\qquad
\mathbf{I}=\mathbf{1},\mathbf{8},
\eeq
where
\begin{align}
\bar g_{\mathbf{I}}(\as)
&={\rm Tr\,}\left[\mathbf{h}^{(0)}\(1+\as\mathbf{h}^{(1)}+\as^2\mathbf{h}^{(2)}+\Ord(\as^3)\)
\mathbf{\Pi_I}\right] \nonumber\\
&=\bar g_{\mathbf{I}}^{(0)}+\as \bar g^{(1)}_{\mathbf{I}}+\as^2\bar g^{(2)}_{\mathbf{I}}+\Ord(\as^3)
\end{align}
are independent of $N$. 

Eq.~\eqref{eq:Stermanfact3} may be cast in a more familiar form, by
reabsorbing the logarithmic terms
in the factor $1+\Ca\as(4m^2/N^2)/\pi$
in a redefinition of the function
$\Gamma_S(\as)$. This also generates
$N$-independent terms, which can be absorbed in a redefinition of
the constants $\bar g^{(i)}_\mathbf{8}$~\cite{Czakon:2009zw}.
Putting everything together, we obtain
\beq
\label{eq:finalratio}
\Sigma^{\rm res}(m^2,\xi,N)
=\hat\sigma^{\sss LO}(m^2,\xi)
\sum_{\mathbf{I=1,8}}
\bar g_{\mathbf{I}}(\as)\exp\left[G_\mathbf{I}(N)\right]
+\Ord\(\frac{1}{N}\)
\eeq
with 
\begin{align}
\label{eq:GI}
&G_\mathbf{I}(N)=\int_0^1dz\,\frac{z^{N-1}-1}{1-z}
\left[2\int_{\muf^2}^{4m^2(1-z)^2}\frac{d\mu^2}{\mu^2}\,
A(\as(\mu^2))+ D_{\mathbf{I}}(\as(4m^2(1-z)^2))\right]
\\
&\qquad D_{\mathbf{1}}(\as)=D(\as)\qquad
\\
&\qquad D_{\mathbf{8}}(\as)=D(\as)+2\Gamma_S(\as)
+2\Ca\beta_0\(\frac{\as}{2\pi}\)^2+\Ord(\as^3).
\end{align}

Finally, recalling the relation Eq.~\eqref{eq:link4} between total
cross section and invariant-mass distribution,  we get
\begin{align}
\hat\sigma(m^2,N)
&=\int_0^1d\xi\,\xi^{N-2}\Sigma(m^2,\xi,N)
\notag\\
&=\hat\sigma^{\sss LO}(m^2,N)
\sum_{\mathbf{I=1,8}}
\bar g_{\mathbf{I}}(\as)\exp\left[G_\mathbf{I}(N)\right]+\Ord\(\frac{1}{N}\)
\label{eq:sigmatotres}
\end{align}
where
\beq
\hat\sigma^{\sss LO}(m^2,N)
=\int_0^1d\xi\,\xi^{N-2}\hat\sigma^{\sss LO}(m^2,\xi),
\label{eq:sigmaires}
\eeq
is the Mellin transform of the Born level total cross section.
By comparing Eq.~\eqref{eq:sigmatotres} with
the definition of the coefficient function Eq.~\eqref{eq:defcoeff}, we
see that
\beq
C(N)=\sum_{\mathbf{I=1,8}}
\bar g_{\mathbf{I}}(\as)\exp\left[G_\mathbf{I}(N)\right]+\Ord\(\frac{1}{N}\)
\label{eq:finalfact}
\eeq
has the same singularity structure as the Mellin-transformed invariant-mass distribution in the $N\to \infty$ limit.

\subsection{Analytic structure of the coefficient function}
\label{sec:total}

We now focus on the coefficient function $C(N)$, implicitly defined in
Eq.~\eqref{eq:defcoeff}. We have shown in the previous section that in
the $N\to \infty$ limit
$C(N)$ has
the same singularity structure as the Mellin-transformed invariant 
mass distribution, whose behaviour in turn is determined by soft-gluon
emissions. 
On the other hand, we 
note that $C\(N\)$ has the same leading singularity in $N=1$ as $\hat{\sigma}
(m^2,N)$, because $\hat{\sigma}^{\sss LO} (m^2,N)$ is  subleading 
in the high-energy regime. Therefore, we can construct an
approximation to $C(N)$ according to the procedure of
Ref.~\cite{Ball:2013bra}, as
\beq
\label{eq:approx}
C_{\rm approx}(N)=C_{\rm soft}(N)+C_{\rm h.e.}(N),
\eeq
where $C_{\rm soft}$ contains the terms predicted by
Sudakov (soft) resummation, and $C_{\mathrm{h.e.}}$ the terms predicted
by BFKL (high-energy) resummation. Explicit expressions for both
components are given in the following Sections.

\section{Large-$N$ contributions} 
\label{sec:largeN}

\subsection{Threshold resummation and analyticity}
\label{sec:softres}

In this subsection, we will extend the procedure outlined in
Ref.~\cite{Ball:2013bra} for the Higgs production cross section to
the coefficient function $C(N)$ of Eq.~\eqref{eq:finalfact}.  The cusp
anomalous dimension $A(\as)$ has been computed up to three
loops~\cite{Vogt:2004mw}, and the colored characteristic anomalous
dimensions $D_{\mathbf{I}}(\as)$ are known completely at two
loops~\cite{Cacciari:2011hy}, allowing us to achieve NNLL accuracy.
Hence, we are able to determine all terms of order $\as^n \ln^m N$,
with $2n-3 \leq m \leq 2n$, in the coefficients of the perturbative
expansions of $C(N)$.  The inclusion of $\as^2$ contribution in the
constant terms $\bar g_{\mathbf{I}}$ enables the extension of our
prediction to $2n-4 \leq m \leq 2n$. We are thus able to predict all
large-$N$ non-vanishing contributions to $C(N)$ up to $\Ord\(\as^2\)$,
and all logarithmically enhanced contributions except the single log
and the constants $\bar g_{\mathbf{I}}^{(3)}$ at $\Ord(\as^3)$. The
single log at $\Ord(\as^3)$, formally a N$^3$LL contribution, is
produced by the third order of the colored characteristic anomalous
dimensions $D_\mathbf{I}$. At this order only the singlet  is fully
known~\cite{Laenen:2005uz}, but we do not know the contributions
to $D^{(3)}_\mathbf{8}$ coming from the N$^3$LO soft anomalous dimension
and soft function. We thus set $D^{(3)}_\mathbf{8}=D^{(3)}_\mathbf{1}$ and $\bar g_{\mathbf{I}}^{(3)}=0$
in the following.  The extra uncertainty induced by this assumption on
our results will be discussed in Sect.~\ref{sec:hadlev} below.

As pointed out in Ref.~\cite{Ball:2013bra,Kramer:1996iq},
the quality of the soft approximation to the full cross section
significantly depends on the choice of subleading terms which are
included in the resummed result. Since the resummation procedure only
fixes the coefficients of logarithmically divergent terms, there is a
certain freedom in defining how the soft approximation is
constructed, by including contributions that are suppressed in the
limit $N\to \infty$.
The idea proposed in Ref.~\cite{Ball:2013bra,Bonvini:2014jma} is to include
subleading terms that are known to be present in the exact calculations,
in order to preserve the known analytic structure at small $N$. 

In principle, the exponent $G_\mathbf{I}(N)$ Eq.~\eqref{eq:GI} is
ill-defined, because the integration range includes the Landau pole of
the strong coupling. This problem is usually avoided by computing the
integral in the large-$N$ limit: the resummed $N$-space result is then
well-defined as a function of $\ln N$. The logarithm, however, has an
unphysical cut at $N=0$, where the cross section should only have
poles. 

However, as pointed out in Ref.~\cite{Ball:2013bra}, if we are only interested
in the expansion of Eq.~\eqref{eq:GI} in powers of $\alpha_s(m^2)$ to
finite order,  the problem of the Landau pole does not arise. The
Mellin transform in Eq.~\eqref{eq:GI} may then be computed exactly, provided
the integrand is understood to be expanded in powers
of $\as(m^2)$ to some finite order. The result is  a function
of $N$ which has the correct logarithmic behaviour at $N\to\infty$, but is
free of  unphysical
cuts on the real negative axis from $N=0$. 

An explicit calculation yields
\beq
\label{eq:Sud}
G_\mathbf{I}(N)=\sum_{n=1}^\infty \as^n\left[\sum_{k=0}^n b^{(A)}_{n,k}
  \mathcal{D}_k(N) +\sum_{k=0}^{n-1}
  b^{(D)}_{n,k,\mathbf{I}}\mathcal{D}_k(N)\right],
\eeq
where $\mathcal{D}_k(N)$ are the Mellin transforms of the distributions
\beq
\mathcal{D}_k(z)=\plus{\frac{\ln^k(1-z)}{1-z}},
\eeq
and the coefficients $b^{(A)}_{n,k}, b^{(D)}_{n,k,\mathbf{I}}$ up to order $n=3$ are
given in the Appendix~\ref{app:largeN}, Eqs.~\eqref{eq:bA}
and~\eqref{eq:bD}, and depend respectively on $A(\as)$ and
$D_{\mathbf{I}}(\as)$ only.  Explicit forms of the functions  
$\mathcal{D}_k(N)$ are given e.g.\ in
Ref.~\cite{Ball:2013bra}; they are seen to only have poles on the real axis.

Following the argument of Ref.~\cite{Ball:2013bra}, we now observe
that the logarithmic enhancement at threshold
arises from the integration  over the
transverse momentum of the emitted gluons, which has the form
\beq
\mathcal{P}_{gg}(z)\int_{\Lambda}^{M\frac{(1-z)}{\sqrt{z}}}\frac{d\kt}{\kt}
=\frac{A_g(z)}{1-z}\left[\ln\frac{1-z}{\sqrt{z}}
  + \ln \frac{M}{\Lambda}\right]
\eeq
where $\mathcal{P}_{gg}(z)$ is the gluon-gluon splitting function for
$z<1$ and $A_g(z)= (1-z) \mathcal{P}_{gg}(z)$, with $A_g(1)=A^{(1)}$
(see Eq.~\eqref{eq:A}).  Thus, it is natural to include subleading
terms in Eq.~\eqref{eq:GI} by restoring the factor of $1/\sqrt{z}$ in
the upper integration bound, and by replacing $A^{(1)}$ by the
expansion of $A_g(z)$ about $z=1$ up to some finite order (keeping the
full expression of $A_g(z)$ is not advisable, because an unphysical
singularity in $z=0$ would appear; see Ref.~\cite{Ball:2013bra}).  

We
therefore replace $G_\mathbf{I}(N)$ with
\begin{align}
\label{eq:GIimproved}
\hat G_\mathbf{I}(N)&=\int_0^1dz\,\frac{z^{N-1}-1}{1-z}
\left\{2\int_{\muf^2}^{\frac{4m^2(1-z)^2}{z}} 
\frac{d\mu^2}{\mu^2}\,z \, A(\as(\mu^2)) 
+D_{\mathbf{I}}\(\as\(\frac{4m^2\(1-z\)^2}{z}\)\)\right\} 
\notag\\
&+c_\mathbf{I},
\end{align}
which differs from $G_\mathbf{I}(N)$ by non-logarithmically enhanced terms.
The factor of $z$ in the first terms arises after expansion
of $A_g(z)$ in powers of $1-z$ to first order; indeed, 
it turns out that
\beq\label{eq:apexp}
A_g(z)
=A_g(1)-(1-z)A_g(1)
+\Ord\((1-z)^2\)=z A_g(1)=z A^{(1)}.
\eeq
An extra power of $z$ simply amounts to a shift of $N$ by one unit in
the Mellin-space result. 
This clearly does not affect the logarithmic behaviour at 
$N\to\infty$.
This extra $z$ factor multiplies also all higher order contributions to
the cusp anomalous dimension $A(\as)$.

The constants $c_\mathbf{I}$ have been introduced by requiring
\beq
\lim_{N\to \infty} \left[\hat G_\mathbf{I}(N)-G_{\mathbf{I}}(N)\right]=0.
\eeq
When expanding in powers of $\as$ as in Eq.~\eqref{eq:Sud},
we can effectively get rid of the constants by simply defining distributions
whose Mellin transform differ by $1/N$ suppressed terms at large $N$, namely
\beq
\label{eq:Dhatk1}
\hat{\mathcal{D}}_k\(z\)=\plus{\frac{\ln^k\(1-z\)}{1-z}}
+\left[\frac{\ln^k\frac{1-z}{\sqrt{z}}}{1-z}-\frac{\ln^k\(1-z\)}{1-z}\right].
\eeq
In this way we find
\beq
\label{eq:Sud5}
\hat G_\mathbf{I}(N)=\sum_{n=1}^\infty 
\as^n\left[\sum_{k=0}^{n} b^{(A)}_{n,k}\hat{\mathcal{D}}_k(N+1)
+\sum_{k=0}^{n-1} b^{(D)}_{n,k,\mathbf{I}}\hat{\mathcal{D}}_k(N)\right],
\eeq
where the coefficients $b^{(A)}_{n,k}$ and $b^{(D)}_{n,k,\mathbf{I}}$
are the same as in Eq.~\eqref{eq:Sud}, and the argument of the Mellin
transform of the distributions associated with the cusp term are
shifted by one. Explicit expressions for the Mellin transforms
$\hat{\mathcal{D}}_k(N)$ are given in Appendix~\ref{app:largeN}.
Note that here, unlike in
Ref.~\cite{Ball:2013bra}, we do not shift   the $D$ terms;
however, the difference is subleading, and the
impact is negligible.

It is important to observe that, unlike in 
the case of Higgs production, the inclusion of the second term in the
expansion Eq.~\eqref{eq:apexp}, does not lead to the full
inclusion of all subdominant contributions of the form $\as^n
N^{-1} \ln^{2n-1} N$. This is because contributions of this order to
$C(N)$ may  arise both
from soft emission, but also from interference of powers suppressed terms
in the leading order cross section.  We will use these
terms (by turning on and off the shift in $N$)
as a way to estimate the uncertainty in our procedure.

\subsection{Coulomb singularities}

In Sect.~\ref{sec:softres} we have studied the large-$N$ terms
originated by soft gluon emission.  In the case of heavy quark pair
production, however, there is another class of contributions which,
order by order in perturbation theory, do not vanish in the large-$N$
limit, altogether unrelated to soft emission. It was pointed out
many years ago~\cite{Fadin:1990wx} that pair interaction dynamics and
bound state effects may be relevant in the threshold regime. This
multiple exchange of ``Coulomb gluons'' between the two heavy quarks
in the final state leads to corrections of order
$(\as/\beta)^m$ with $\beta=\sqrt{1-\rho}$, which are
usually referred to as Coulomb singularities.  For
$m$ large enough, such contributions may compete with, and even
dominate over, Sudakov logarithms in the threshold limit.  It turns
out, however, that these contributions can be resummed to all orders
(see for example Ref.~\cite{Frixione:1997ma} and references therein),
and the result of resummation vanishes in the large-$N$
limit. Nevertheless, these contributions must be taken into account in
the absolute threshold limit $\rho\to 1$.

Coulomb terms have been
studied in Refs.~\cite{Fadin:1990wx,Frixione:1997ma,Beneke:2010da,
  Bonciani:1998vc,Cacciari:2011hy,Beneke:1999qg}.  In particular, it
was shown in Ref.~\cite{Beneke:2010da} that Coulomb singularities and
soft singularities factorize in Mellin space in the $N\to \infty$ limit and in
the singlet-octet basis:
\beq
\label{eq:finalfact0}
C(N)=\sum_{\mathbf{I=1,8}} \bar g_{\mathbf{I}}(\as)J_{\mathbf{I}}(N,\as)
\exp\left[G_\mathbf{I}(N)\right]+\Ord\(\frac{1}{N}\),
\eeq
where the factor
\beq
J_{\mathbf{I}}(N,\as)=1+J_{\mathbf{I}}^{(1)}(N)\as
+J_{\mathbf{I}}^{(2)}(N)\as^2+J_{\mathbf{I}}^{(3)}(N)\as^3+\Ord(\as^4)
\eeq
contains Coulomb terms. The coefficients $J_{\mathbf{I}}^{(1)}(N)$ and 
$J_{\mathbf{I}}^{(2)}(N)$ can be obtained by matching with exact fixed-order 
calculations~\cite{Beneke:2009ye}, while $J_{\mathbf{I}}^{(3)}(N)$ is
unknown.

\begin{figure} [tb]
\begin{center}
\includegraphics[width=0.495\textwidth]{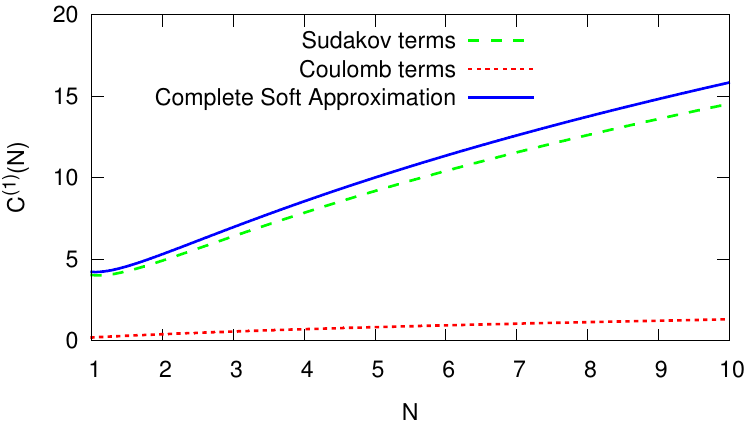}
\includegraphics[width=0.495\textwidth]{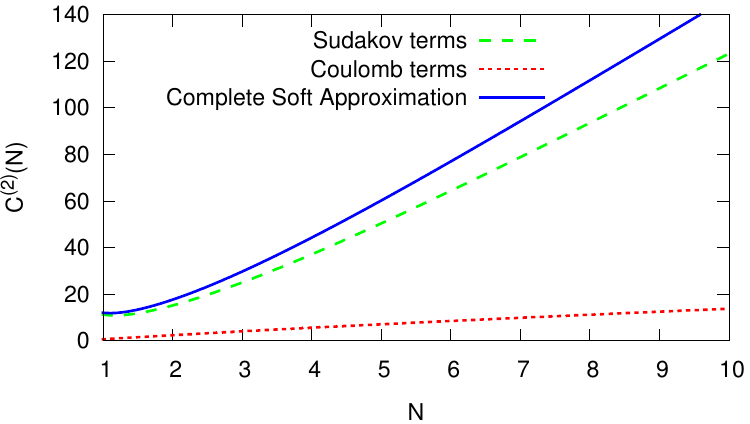}
\caption{The contributions to $C(N)$ from 
  soft gluon emission
  $\exp\left[G_{\mathbf{I}}(N,\as)\right]$ (Sudakov terms) and
  Coulomb terms $J_{\mathbf{I}}(N,\as)$ (Coulomb terms),
  at NLO (left) and NNLO (right). The combined effect
  Eq.~\eqref{eq:finalfact0}  is also shown (Complete soft
  approximation).}
\label{fig:confrontoCoulombSoft1}
\end{center}
\end{figure}

It can be shown that the effect of Coulomb terms on the physical cross
section is very small.  In Ref.~\cite{Cacciari:2011hy} the impact of
the contribution $J_{\mathbf{I}}^{(2)}(N)$ on the hadronic cross
section was estimated to be about $0.5\%$ of total NNLO correction,
and even less at higher orders. Similar conclusion can be drawn
from inspection of Fig.~\ref{fig:confrontoCoulombSoft1}, where we
compare, in $N$ space, the coefficient function Eq.~\eqref{eq:finalfact0}
with and without the Coulomb factors $J_\mathbf{I}(N,\as)$,
both at NLO and NNLO.
The effect at $N\sim 2$, which is the relevant region for Mellin inversion
(see Section~\ref{sect:saddle}), is very small.

\begin{figure} [tb]
\begin{center}
\includegraphics[width=0.9\textwidth]{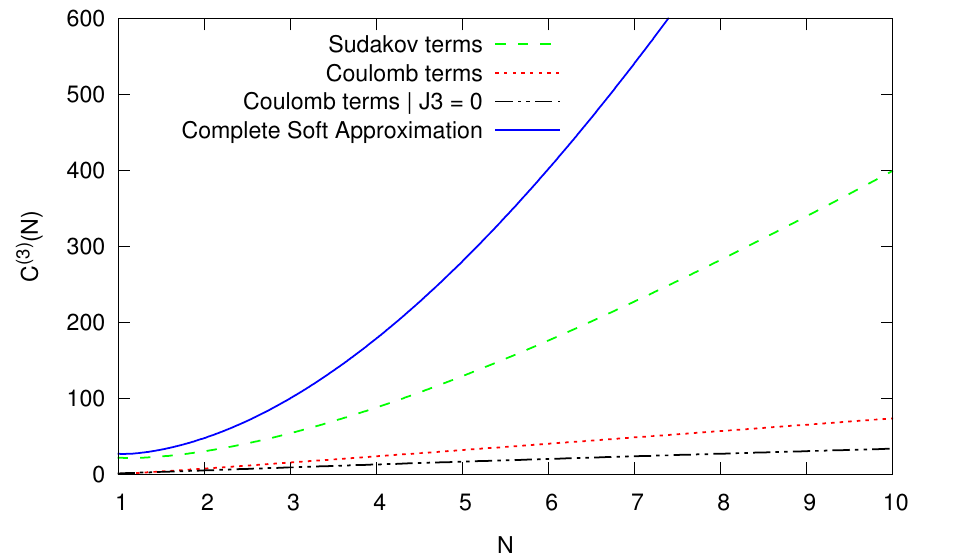}
\caption{Same as Fig.~\ref{fig:confrontoCoulombSoft1} but at
  N$^3$LO. The two different Coulomb curves
  differ by the inclusion of the pNRQCD estimate of  $J_{\mathbf{I}}^{\(3\)}$
  (see text).}
\label{fig:confrontoCoulombSoft2}
\end{center}  
\end{figure}

A complete resummed expression of Coulomb contributions has been obtained
in the context of potential
non-relativistic QCD (pNRQCD)~\cite{Brambilla:1999xf, Beneke:2011mq,
  Pineda:2006ri}. This calculation can be used to test whether
the pattern observed at NLO and NNLO continues at higher orders.
In Fig.~\ref{fig:confrontoCoulombSoft2} we present the same comparison
as in Fig.~\ref{fig:confrontoCoulombSoft1} at N$^3$LO, with $J_\mathbf{I}^{(3)}(N)$
extracted from pNRQCD calculations~\cite{Brambilla:1999xf, Beneke:2011mq,
  Pineda:2006ri}. The effect of the inclusion
of the $\Ord(\as^3)$ term in $J_\mathbf{I}(N,\as)$ as predicted by pNRQCD computations is indeed very small.
We will include $J_\mathbf{I}^{(3)}(N)$ as computed by pNRQCD methods
in our results, but these observations  show that our final results
are not significantly affected by it.
An explicit form for all the Coulomb
corrections is given in Appendix~\ref{app:largeN}, Eq.~\eqref{eq:J} 
(see also Ref.~\cite{Muselli:tesi}).

\subsection{Final prescription for the soft contribution}
\label{sec:finalsoft}

We conclude this section by giving our final prediction for the
soft-emission contribution to the total cross section
to N$^3$LO.
We define
\begin{align}
C_{\text{A-soft}_0}(N)&=\sum_{\mathbf{I=1,8}}
\bar{g}_\mathbf{I}(\as)J_{\mathbf{I}}(N,\as)
\notag\\ 
&\exp\left\{\sum_{n=1}^\infty \as^n\left[\sum_{k=0}^n
  b^{(A)}_{n,k} \hat{\mathcal{D}}_k\(N\)+\sum_{k=0}^{n-1}
  b^{(D)}_{n,k,\mathbf{I}}\hat{\mathcal{D}}_k(N)\right]\right\};
\label{eq:soft0final}
\\
C_{\text{A-soft}_1}(N)&=\sum_{\mathbf{I=1,8}}
\bar{g}_{\mathbf{I}}(\as)
J_\mathbf{I}(N,\as)
\notag\\
&\exp\left\{\sum_{n=1}^\infty \as^n\left[\sum_{k=0}^n
b^{(A)}_{n,k} \hat{\mathcal{D}}_k(N+1)+\sum_{k=0}^{n-1}
b^{(D)}_{n,k,\mathbf{I}}\hat{\mathcal{D}}_k(N)\right]\right\} 
\label{eq:soft1final}
\end{align}
where we have made explicit the  dependence on $\as=\as(m^2)$ and it is
understood that  the exponentials should be expanded in powers of
$\as$ up to order $\as^3$.
We will take the average between the two approximations,
\beq\label{eq:softfinal}
C_{\text{A-soft}}(N)=\frac{1}{2}
\left[C_{\text{A-soft}_1}(N)+C_{\text{A-soft}_0}(N)\right]
\eeq
as the central value of our prediction, and the difference
\beq
\label{eq:softfinalerr}
\Delta^{\mathrm{soft}}(N)=
\left|C_{\text{A-soft}_1}(N)-C_{\text{A-soft}_0}(N)\right|
\eeq
as an estimate of the uncertainty of our procedure.
Hence, our soft approximation to the coefficient function will be finally
given by
\beq\label{eq:softapp}
C(N)=C_{\text{A-soft}}\pm\frac{\Delta^{\rm soft}}{2}.
\eeq

We also consider the result which corresponds to the standard
resummation, as given to NLL in Ref.~\cite{Bonciani:1998vc}, and
extended to NNLL in
Ref.~\cite{Czakon:2013goa} (and in the associate public
code~\texttt{top++}), which is expressed as a function of 
positive powers of $\ln N$, which we will call
$N$-soft, following Ref.~\cite{Ball:2013bra}. This is given by
\begin{align}
C_{N\text{-soft}}(N)&=\sum_{\mathbf{I=1,8}}
g_\mathbf{I}(\as)J_{\mathbf{I}}(N,\as)
\notag\\ 
&\exp\left\{\sum_{n=1}^\infty \as^n\left[\sum_{k=0}^n
  b^{(A)}_{n,k} \mathcal{D}^\text{log}_k\(N\)+\sum_{k=0}^{n-1}
  b^{(D)}_{n,k,\mathbf{I}}\mathcal{D}^\text{log}_k(N)\right]\right\},
\label{eq:Nsoft}
\end{align}
where the functions $\mathcal{D}^\text{log}_k(N)$ are the large $N$ limit
of $\mathcal{D}_k(N)$ written in terms of positive powers of $\ln N$, 
\emph{excluding} constants
(for an explicit definition, see Ref.~\cite{Ball:2013bra}), i.e. such that
\begin{equation}
\mathcal{D}_k(N)=\mathcal{D}^\text{log}_k(N)+d_k+\Ord\left(\frac{1}{N}\right),
\end{equation}
thereby leading to
\begin{equation} \label{eq:g0vsg0bar}
g_\mathbf{I}(\as)=\bar{g}_\mathbf{I}(\as)\exp\left[ \sum_{n=1}^3\as^n 
\left(\sum_{k=0}^n b^{(A)}_{n,k}d_k + \sum_{k=0}^{n-1} b^{(D)}_{n,k,\mathbf{I}}d_k 
\right) \right].
\end{equation}

\section{Small-$N$ contributions}
\label{sec:smallN}

In order to extract the leading small-$N$ singularity of the partonic
cross section, we will use the so-called high-energy or $\kt$
factorization technique, first described in Ref.~\cite{Catani:1990eg}
for the total cross section, and more recently extended to rapidity
distributions~\cite{Ball:1999sh}.  We follow the resummation procedure
developed by Altarelli, Ball and Forte (ABF)~\cite{Altarelli:2008aj}.
For a detailed derivation of resummation, which affects both
coefficient functions and evolution of the parton densities, we refer
the reader to the original literature
(e.g.~\cite{Catani:1990eg,Catani:1993ww,Altarelli:2008aj,Altarelli:2005ni});
here we only summarize the main results which  are relevant
for our discussion.

In the $\kt$ factorization formalism, small-$N$ singularities are
obtained by computing the leading-order partonic cross section for the
relevant process, with off-shell incoming gluons. We therefore define
an off-shell partonic cross section $\frac{\as^2}{m^2}\sigg_\text{off-shell}(\rho,\xi_1,\xi_2)$, which is a function of the scaling
variable of the process $\rho$, and of the transverse momenta of the
two off-shell gluons: $\xi_i={k^2_{\sss T}}_i/m^2$. Resummed results can
be obtained through the determination of the so-called impact factor
\begin{align}
h(N,M_1,M_2,m^2,\as)&= M_1 M_2 R(M_1)R(M_2)\(\frac{m^2}{\muf^2}\)^{M_1+M_2}
\notag \\
&\times \int_0^1d\rho\,\rho^{N-2}\int_0^\infty d\xi_1\,
\xi_1^{M_1-1}\int_0^\infty d\xi_1\,\xi_1^{M_2-1}
\frac{\as^2}{m^2}\hat\sigma_\text{off-shell}(\rho,\xi_1,\xi_2).
\label{eq:impact}
\end{align}
The prefactor $R(M_1)R(M_2)$ accounts for factorization scheme
dependence~\cite{Catani:1993ww}; in the \MSbar\ scheme
\beq
R(M)=1+\frac{8}{3}\zeta_3 M^3+\Ord(M^4).
\eeq
The impact factor Eq.~\eqref{eq:impact} for the production of a heavy quark
pair was computed in Ref.~\cite{Ball:2001pq}.  Here, we are interested in its 
expansions in powers of $M_1,M_2$, in the vicinity of $N=1$:
\beq \label{eq:high1}
h(N,M_1,M_2,m^2,\as)
=\frac{\as^2}{m^2}\sum_{i_1,i_2}^{\infty} h_{i_1,i_2}(\muf^2) M_1^{i_1} M_2^{i_2} + \Ord(N-1).
\eeq
The coefficients $h_{i_1,i_2}$ relevant at N$^3$LO can be easily obtained by performing the expansions of the previous formula and they are given in the Appendix~\ref{app:smallN}, Eq.~\eqref{eq:himpact}.
Following Refs.~\cite{Ball:2007ra,Altarelli:2008aj}, the resummation is performed by identifying the Mellin variable $M_i$ with the resummed DGLAP anomalous dimension (together with sub-leading running-coupling effects):
\begin{equation}
M_i^k =\left[{\gammapres}^k\right],
\end{equation}
where the right hand side is recursively defined by
\beq
\label{eq:gammapsquare}
\left[{\gammapres}^{k+1}\right]=\gammapres\(1+k\frac{\dotgammapres}
{{\gammapres}^2}\)
\left[{\gammapres}^k\right],\quad \left[\gammapres\right]=\gammapres 
\eeq
with
\beq
\label{eq:gammappunto}
\dotgammapres=-\beta_0\as^2\frac{\partial}{\partial\as}\gammapres.
\eeq

By expanding the anomalous dimension to fixed perturbative order
\beq
\label{eq:gammapiu}
\gamma^{\rm\sss +}=\as \gamma^{(0)}+ \as^2 \gamma^{(1)}+\as^3 \gamma^{(2)}+ \Ord(\as^4),
\eeq
we have all the ingredients to
construct our $N\to 1$ approximation, which is simply found by
substituting the expansion~\eqref{eq:gammapiu} in
Eq.~\eqref{eq:gammapsquare}, and then in Eq.~\eqref{eq:high1}. The
explicit form of the anomalous dimensions in the small-$N$ limit to the
order relevant in our discussion is given in
Appendix~\ref{app:smallN}, Eq.~\eqref{eq:gamma0},~\eqref{eq:gamma1}
and~\eqref{eq:gamma2}. Now we define a resummed coefficient function
by factoring out the leading order contribution, evaluated in the
high-energy limit, i.e. $N=1$
\begin{align}
C_{\sss ABF}(N)
&=\frac{\as^2}{m^2}\frac{1}{\sigg^{\sss LO}(m^2,1)}
\(\sum_{i_1,i_2=0}^{\infty}h_{i_1,i_2}
\left[\gammapres\raisebox{1.6ex}{$\scriptstyle i_1$}\right]
\left[\gammapres\raisebox{1.6ex}{$\scriptstyle i_2$}\right]\)-1
\notag\\
&=2\bar{h}_{1,0}\gamma^{\(0\)}
+\as^2\left[\(2\bar{h}_{2,0}+\bar{h}_{1,1}\)
\gamma^{(0)}\raisebox{1.6ex}{$\scriptstyle 2$}
-2\bar{h}_{2,0}\beta_0\gamma^{\(0\)}+2\bar{h}_{1,0}\gamma^{(1)}\right]
\notag\\
&+\as^3\Big[\(\bar{h}_{3,0}+\bar{h}_{2,1}\)
2\gamma^{(0)}\raisebox{1.6ex}{$\scriptstyle 3$}
-(3\bar{h}_{3,0}+\bar{h}_{2,1})
2\beta_0\gamma^{(0)}\raisebox{1.6ex}{$\scriptstyle 2$}
+4\bar{h}_{3,0}\beta_0^2\gamma^{(0)}
\nonumber\\
&\qquad+(2\bar{h}_{2,0}+\bar{h}_{1,1})2\gamma^{(0)}\gamma^{(1)}
-4\bar{h}_{2,0}\beta_0\gamma^{(1)}
+2\bar{h}_{1,0}\gamma^{(2)}\Big]+\Ord(\as^4),
\label{eq:ABF}
\end{align}
where, in the second equality, we use the fact that 
\beq
\sigg^{\sss LO} (m^2,1)=\frac{\as^2}{m^2}\,h_{0,0},
\eeq
and we define
\beq
\bar{h}_{i_1,i_2}=\frac{h_{i_1,i_2}}{h_{0,0}}.
\eeq
We have omitted the dependence of the coefficients on
$\muf^2$ for simplicity.  

Since we are going to combine the small-$N$ approximation with 
the large-$N$ approximation to obtain an estimate of the full coefficient 
function,
we require that the two limiting behaviors do not interfere with each other.
In particular, we require that the high-energy contribution vanishes when
$N\to\infty$. Manifestly, Eq.~\eqref{eq:ABF} does not fulfil this requirement,
because of the presence of a constant contribution
in $\gamma^{(0)}$ (see Eq.~\eqref{eq:gamma0} in Appendix~\ref{app:smallN}), which propagates in $C_{\sss ABF}(N)$ to all
orders in $\as$.  Therefore, following Ref.~\cite{Ball:2013bra},
we replace Eq.~\eqref{eq:ABF} with a modified version
of the small-$N$ approximation to the partonic cross section,
which has the same leading
singularity in $N=1$ but vanishes as $N\to\infty$. The modified version is
given by
\beq
\label{eq:ABFsub}
C_{\sss\text{ABF-sub}}(N)
=C_{\sss ABF}(N)-2C_{\sss ABF}(N+1)
+C_{\sss ABF}(N+2).
\eeq
The subtraction simply introduces subleading
singularities in $N=0$ and $N=-1$, but now
\beq
\lim_{N\to\infty}C_{\sss\text{ABF-sub}}(N)=0
\eeq
to all orders in $\as$. As pointed in Ref.~\cite{Ball:2013bra},
this choice for the subtraction is
a compromise between the contrasting goals of not changing the
small-$N$ singularities structure and of damping strongly enough as $N$
increases. In momentum space the subtraction of Eq.~\eqref{eq:ABFsub}
corresponds to damping the $\rho\to 1$ behaviour of the partonic cross
section through a multiplicative factor $(1-\rho)^2$.

One additional modification is needed to define
our prediction for the cross section in the high-energy regime.
We note that the anomalous dimensions vanishes at
$N=2$ due to
momentum conservation. This implies that the small-$N$ approximation
Eq.~\eqref{eq:ABF} vanishes in
$N=2$. This value of $N$ marks the transition between the small-$N$
approximation (not accurate if $N\gtrsim 2$) and the large-$N$
approximation (not accurate when $N\lesssim 2$).

However, Eq.~\eqref{eq:ABF}, differs from $0$ in $N=2$
because the small-$N$ limit of the anomalous dimension $\gamma^+$,
Eq.~\eqref{eq:gammapiu}, contains only the leading and the
next-to-leading singularities in $N=1$, and not a full fixed order
expression. Momentum conservation can be enforced~\cite{Altarelli:2008aj}
directly in Eq.~\eqref{eq:ABF}, by adding to
$C_{\sss ABF}(N)$ a function $f_{\rm mom}\propto1/N$. Following
Ref.~\cite{Ball:2013bra}, we construct our small-$N$ momentum-conserving approximation as
\beq
\label{eq:HE}
C_{\rm h.e}(N)
=C_{\sss\text{ABF-sub}}(N)
-\frac{4!k_{\rm mom}(\as)}{N(N+1)(N+2)},
\eeq
where the constant $k_{\rm mom}(\as)$ is fixed by requiring
that $C_{\rm h.e}(2)=0$ to all orders in the perturbative
expansion. The enforcement of momentum conservation in our small-$N$
approximation allows us to estimate the contribution
of subleading poles in $N=0,-1,-2,\dots$, which are not
controlled by the high-energy approximation. 
Indeed, the full partonic cross
section does not vanish at $N=2$:  only the contribution to it
driven by hard radiation from external legs does. Hence we can estimate
subleading small-$N$ effects by allowing the partonic cross section to
deviate slightly from zero at $N=2$ due to  contributions from
subleading poles. We thus fix $k_{\rm mom}$ to be
\beq
\label{eq:kmom}
k_{\rm mom}(\as)=C_{\sss\text{ABF-sub}}(2)\pm \Delta k_{\rm mom}(\as),
\eeq
and we take the variation $\Delta k_{\rm mom}(\as)$ to reach as its
maximum value $15\%$ of the size of soft contribution
$C_{\text{A-soft}}(N)$, Eq.~\eqref{eq:softfinal} at
$N=2$.  This is a somewhat more conservative estimate of the
uncertainty in comparison to the corresponding one adopted in
Ref.~\cite{Ball:2013bra}, due to the fact that we expect the
contribution from subleading
poles to be potentially somewhat larger in this case, based on the
behaviour of known orders, and possibly also due to the issues
discussed in Sect.~\ref{sect:sings}.
Namely, we choose in Eq.~\eqref{eq:kmom},
\beq\label{eq:deltak}
\Delta k_{\rm mom}(\as)=0.15\times C_{\text{A-soft}}(2).
\eeq
In conclusion, this means that the small $N$ contribution, rather than
being completely switched off at $N=2$, is small at this point, and
gets switched off somewhere in its vicinity.

\section{Approximate cross section up to N$^3$LO}
\label{sec:approx}

We are now ready to present our results for top pair production
cross section at N$^3$LO, by using Eq.~\eqref{eq:approx} to combine the
large-$N$ terms Eq.~(\ref{eq:softapp}) and the small-$N$ terms 
Eq.~(\ref{eq:HE}). We first recall how $N$-space parton-level
results can be related to physical hadron-level results using the
saddle point methods, and then we  present the parton and hadron
level results in turn.

\subsection{Saddle point approximation of Mellin inversion integrals}
\label{sect:saddle}

The physical hadronic cross section can be related to the underlying
partonic cross section by  viewing the former as the inverse Mellin
transform of its factorized expression Eq.~(\ref{eq:melfac})
\beq
\sigma(m^2,\rho_h)=\frac{1}{2\pi i}\int_{c-i\infty}^{c+i\infty} dN\,
\rho_h^{-N+1}\mathscr{L}(N)\hat{\sigma}(m^2,N)
=\frac{1}{2\pi i}\int_{c-i\infty}^{c+i\infty} dN\,\exp[f(N)],
\eeq
where $c$ is to the right of all singularities of the integrand. It can be shown
on general grounds that $f(N)$ has a unique minimum at $N=N_0$ 
on the real $N$ axis, which allows us to estimate the integral
by the saddle-point technique.

The position $N_0$ of the saddle typically depends very weakly 
on the partonic cross section, 
and is mainly determined by the parton luminosity; the saddle-point
approximation, with the inclusion of quadratic fluctuations around the
saddle, turns out to be generally quite accurate~\cite{Bonvini:2014qga}.
It is then possible to infer properties of the hadronic cross section
from the behaviour of the partonic cross section at the value of $N$
which corresponds to the saddle point for given hadronic kinematics.

\begin{figure}[t]
  \begin{center}
    \includegraphics[width=0.8\textwidth,page=2]{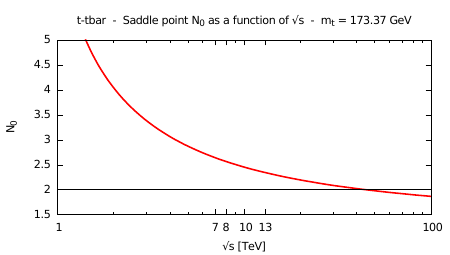}
  \end{center}
\caption{Position of the saddle point as a function
of $\sqrt{s}$ for fixed top quark mass.}\label{fig:saddle}
\end{figure}

The position of the saddle point
$N_0$ as a function of the collider energy $\sqrt{s}$, for $m=m_t=173.37$~GeV, computed
using NNPDF 3.0~\cite{Ball:2014uwa} parton distributions, is shown in Fig.~\ref{fig:saddle}.   
The value of the saddle turns out to be around $N=2.5$
at LHC energies. The value of $N_0$
is a very slowly decreasing function of the total energy.

\subsection{Parton-level results}
\label{sec:part}

We can now combine the results of Sections~\ref{sec:largeN} 
and~\ref{sec:smallN} to obtain an estimate of the coefficients of
the perturbative expansion for the total production cross section 
of heavy quark pairs.
We first test our procedure against exact fixed-order
calculations, which are available at
NLO~\cite{Nason:1987xz,Czakon:2008ii} and NNLO~\cite{Czakon:2013goa}.
At NLO we specifically compare to the fit to the analytic result of
Ref.~\cite{Czakon:2008ii} presented in
Ref.~\cite{Aliev:2010zk} .
\begin{figure} [htb]
\begin{center}
\includegraphics[width=0.495\textwidth]{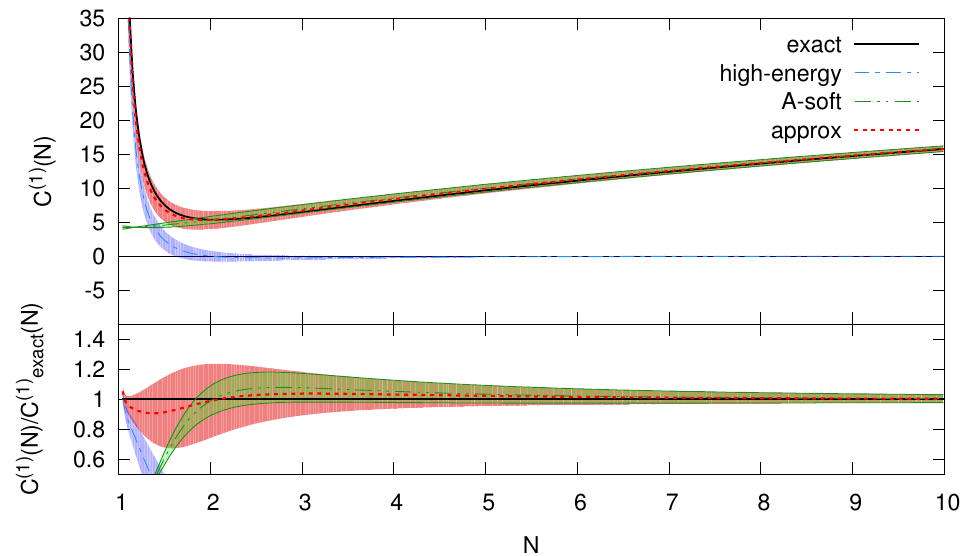}
\includegraphics[width=0.495\textwidth]{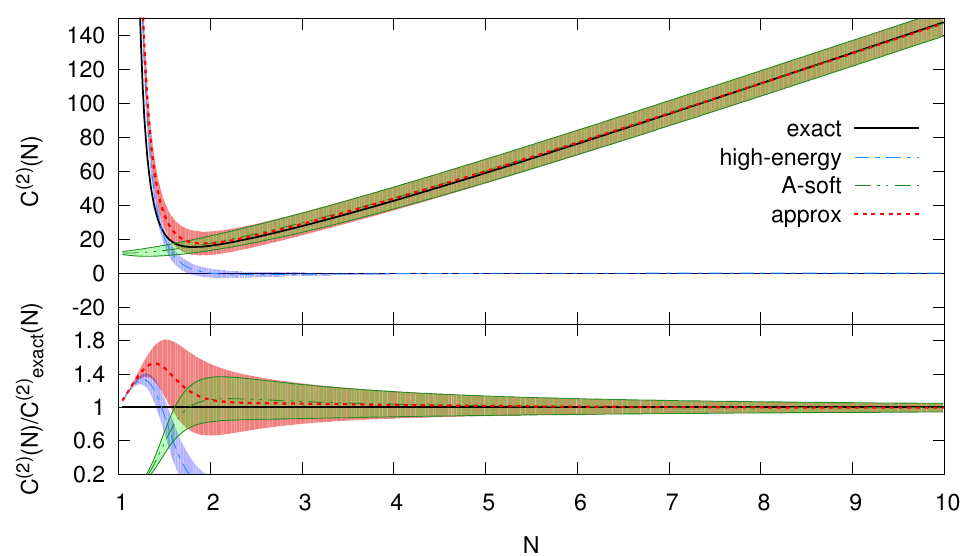}
\caption{Comparison between exact results and our
  approximation at NLO (left) and NNLO (right) for the Mellin-space
  coefficient function $C(N)$; the large-$N$
  contribution (A-soft) Eq.(\ref{eq:softapp}), the small-$N$ contribution
  (high-energy) Eq.(\ref{eq:HE}), and
  the combined approximation of Eq.~\eqref{eq:approx} (approx) are shown. The
  bottom plots show the ratio of the approximate to
  the exact result.}
\label{fig:NLONNLO}
\end{center}
\end{figure}
In Fig.~\ref{fig:NLONNLO} we show our approximation to the
 NLO and NNLO in Mellin
space (labeled approx), compared to the exact results of
Ref.~\cite{Aliev:2010zk,Czakon:2013goa}. 
The large-$N$ contribution, $C_{\text{A-soft}}(N)$
Eq.~\eqref{eq:softfinal} (labeled as soft), and the small-$N$
contribution, $C_{\rm h.e}(N)$ Eq.~\eqref{eq:HE} (labeled as
high-energy), are also  shown. The renormalization and
factorization scales are taken to be equal to the heavy quark mass
$m$.
The uncertainty on our prediction (red band) 
is obtained as the envelope of the uncertainty Eq.~(\ref{eq:softapp})
on the soft terms (green band), and the uncertainty
Eqs.~\eqref{eq:kmom},~\eqref{eq:deltak} on the  high-energy terms (blue band).

The agreement is excellent at NLO in the whole range displayed in
Fig.~\ref{fig:NLONNLO}.  At NNLO there is a slight discrepancy in the
region between $1<N<1.6$. This is due to the fact we are only
including the LL contribution in this region, while it was noted in
Ref.~\cite{Czakon:2013goa, Czakon:2008ii} that NLL contributions close to $N=1$
for top pair production are sizable. This region of $N$, however,
would only be relevant for very high energy colliders ($\sqrt{s}\gsim
100$ TeV). 

At the $N$ values which are relevant for LHC energies, 
the agreement between the approximate and exact results is excellent;
for lower collider energy, as the saddle point moves towards larger
values, the uncertainty on the approximate prediction is smaller.

\begin{figure} [tb]
  \begin{center}
\includegraphics[width=0.80\textwidth]{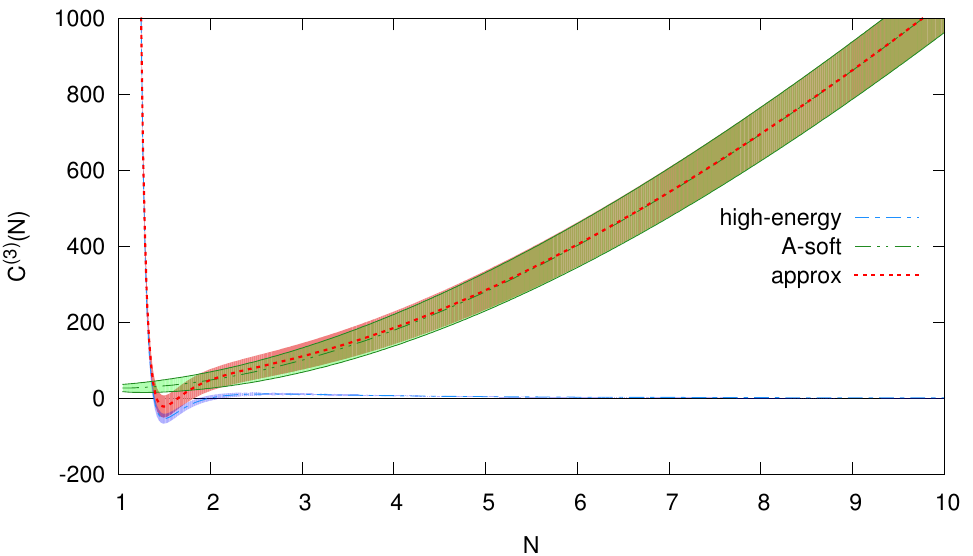}
\caption{Same as Fig.~\ref{fig:NLONNLO}, but at
   N$^3$LO.}
\label{fig:NNNLO}
\end{center}
\end{figure}

We now turn to the N$^3$LO contribution.
We recall that the constants $\bar g_{\mathbf{I}}^{(3)}$, together with
the difference between coefficients $D_{\mathbf{8}}^{(3)}$ and
$D_{\mathbf{1}}^{(3)}$, are not known. They will be set to zero for
the time being;  the impact of this missing
information on our prediction for the physical cross section will be
discussed in Section.~\ref{sec:hadrocross} below.
The function $C_{\rm approx}^{(3)}(N)$ is plotted in
Fig.~\ref{fig:NNNLO}, together with the soft 
and the high-energy contributions that go into it according to
Eq.~(\ref{eq:approx}).

\subsection{Hadron-level results}
\label{sec:hadlev}

We come now to  the hadronic  cross sections. It is convenient
to define the gluon-channel $K$-factors
\beq
\label{eq:Kfactor}
K_{gg}(m^2,\rho_h)
=\frac{\sigma_{gg}(m^2,\rho_h)}{\sigma_{gg}^{\sss LO}(m^2,\rho_h)}
=1+\as K_{gg}^{(1)}+\as^2 K_{gg}^{(2)}+\as^3 K_{gg}^{(3)}+\Ord(\as^4),
\eeq
where $\as=\as(m^2)$, $\sigma_{gg}(m^2,\rho_h)$ is the contribution to
the hadron-level cross section Eq.~\eqref{eq:hqpcross} from the gluon-gluon
subprocess, and $\sigma_{gg}^{\mathrm{LO}}(m^2,\rho_h)$ the corresponding
leading-order approximation. 
All results will be obtained  using the partonic cross sections of
Sect.~\ref{sec:part}, with  factorization and renormalization scale 
$\mur=\muf=m$, and NNPDF $3.0$ NNLO parton
distribution functions~\cite{Ball:2014uwa}, with $\as(M^2_Z)=0.118$
and $\Nf=5$. Scale uncertainties on our final results will be
discussed in Sect.~\ref{sec:hadrocross} below.

\begin{figure} [!htbp]
\centering
\includegraphics[width=0.80\textwidth]{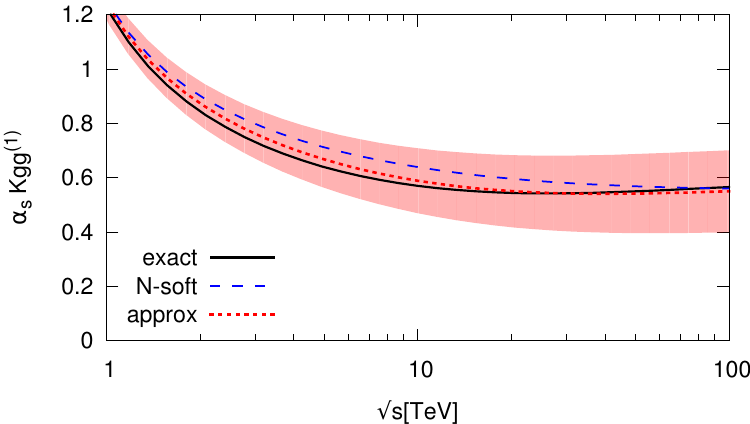}
\caption{ Comparison between the  exact result and our
  approximation for the
NLO contribution to the $K$-factor Eq.~\eqref{eq:Kfactor} from
  the gluon channel, plotted as a function of the collider energy
  $\sqrt{s}$. The result obtained expanding the standard 
  NNLL resummation ($N$-soft) is also shown.}
\label{fig:Kfactor1}
\end{figure}
\begin{figure} [!htbp]
\centering
\includegraphics[width=0.80\textwidth]{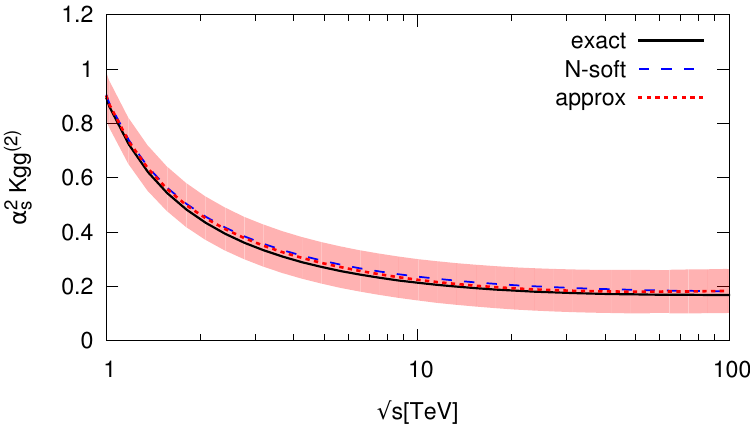}
\caption{Same as Fig.~\ref{fig:Kfactor1} at NNLO.}
\label{fig:Kfactor2}
\end{figure}
\begin{figure} [!htbp]
\centering
\includegraphics[width=0.80\textwidth]{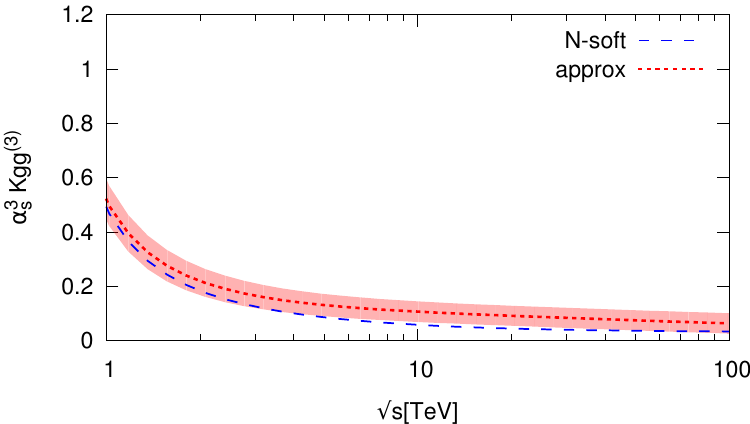}
\caption{ Same as Fig.~\ref{fig:Kfactor1} at N$^3$LO, where the exact
  result is not known.}
\label{fig:Kfactor3}
\end{figure}

The NLO and NNLO $K$-factors at a $pp$ collider are shown in
Figs.~\ref{fig:Kfactor1} and~\ref{fig:Kfactor2}, respectively, as
functions of the center-of-mass energy $\sqrt{s}$, and compared to
the exact results of Ref.~\cite{Aliev:2010zk,Czakon:2013goa}. 
We also show the values obtained by expanding out to the given order
the standard resummed result  of
Refs.~\cite{Bonciani:1998vc,Czakon:2013goa}, i.e.\ using
the $N$-soft approximation Eq.~\eqref{eq:Nsoft}.
The main result of this work, namely the N$^3$LO contribution to the
$K$-factor in the gluon-gluon channel as a function of the collider
energy is shown in Fig.~\ref{fig:Kfactor3}.
 Numerical results for the $K$-factors at LHC energies are
collected in Tab.~\ref{tab:results}.

\begin{table}[t]
\centering 
\begin{tabular} {cc|lll}
&& $\as(m^2) K_{gg}^{(1)}$ & $\as^2(m^2) K_{gg}^{(2)}$ & $\as^3(m^2) K_{gg}^{(3)}$ \\
\midrule
\parbox[t]{2mm}{\multirow{3}{*}{\rotatebox[origin=c]{90}{$7$ TeV}}}
& \textrm{exact}  & 0.599 & 0.237 &  \\
& \textrm{approx} & $0.623\pm 0.112$ & $0.250\pm 0.076$ & $0.098\pm 0.038$ \\
& $N$\textrm{-soft}   & 0.672 & 0.260 & 0.056 \\
\hline
\parbox[t]{2mm}{\multirow{3}{*}{\rotatebox[origin=c]{90}{$8$ TeV}}}
& \textrm{exact}  & 0.587 & 0.227 &  \\
& \textrm{approx} & $0.609\pm 0.115$ & $0.239\pm 0.076$ & $0.094\pm 0.037$ \\
& $N$\textrm{-soft}   & 0.658 & 0.250 & 0.052 \\
\hline
\parbox[t]{2mm}{\multirow{3}{*}{\rotatebox[origin=c]{90}{$13$ TeV}}}
& \textrm{exact}  & 0.555 & 0.199 &  \\
& \textrm{approx} & $0.569\pm 0.125$ & $0.209\pm 0.077$ & $0.083\pm 0.036$ \\
& $N$\textrm{-soft}   & 0.619 & 0.221 & 0.040 \\
\hline
\parbox[t]{2mm}{\multirow{3}{*}{\rotatebox[origin=c]{90}{$14$ TeV}}}
& \textrm{exact}  & 0.552 & 0.196 &  \\
& \textrm{approx} & $0.565\pm 0.126$ & $0.205\pm 0.077$ & $0.082\pm 0.036$ \\
& $N$\textrm{-soft}   & 0.614 & 0.217 & 0.038 \\
\hline
\end{tabular}
\caption{The NLO, NNLO, N$^3$LO contributions the gluon
  channel $K$-factor at the  LHC ($7$ TeV, $8$ TeV, $13$ TeV, $14$ TeV).}
\label{tab:results}
\end{table}

We note that our approximation reproduces the
exact result within the estimated uncertainty both at NLO and at NNLO, 
in the whole energy range displayed in the plots.
In comparison to the case of Higgs production, the uncertainty is larger, both
because threshold resummation is known to a
lower accuracy, and also because, as already mentioned, 
we have less control over $\frac{\ln^k N}{N}$ terms.
Our result is seen to differ by varying amounts at each order from the
$N$-soft one, simply obtained by expanding out the resummed result.
At NLO and NNLO the origin of this difference is twofold: first, our
approximation also includes the high-energy contribution and second, the
functional form of the contribution obtained by expanding out the
$N$-soft result differs from that adopted in our approximation, as explained in
Sect.~\ref{sec:finalsoft}.
At N$^3$LO  two further differences are, first,
that our result includes the single logarithmic term, 
even if its coefficient is only partially known, which
is absent in a NNLL resummation. However, the numerical impact of this
contribution is small. Second,    that  
the function of $\alpha_s$ which multiplies the resummed
exponent is given by $ g_\mathbf{I}(\as)$  in the $N$-soft resummed
result, and by $\bar g_\mathbf{I}(\as)$ when constructing our
approximation, the relation between the two being given by
Eq.~(\ref{eq:g0vsg0bar}). We will specifically discuss the impact of
this choice in the next Section (see in particular
Tab.~\ref{tab:compH3}).

\subsection{N$^3$LO top pair production cross section at LHC}
\label{sec:hadrocross}

We collect now final results for 
top pair
production at the  LHC energies and its uncertainty.
 Our prediction for the total cross section with 
 $\muf=\mur=m$ is 
\begin{subequations}
\label{eq:hqpapprox}
\begin{align}
\textrm{LHC$7$:}\; \sigma^{\mathrm{N^3LO}}_{\mathrm{approx}}\(\rho_h,m^2\)
&=\(177.43\pm 2.99 +0.10\, \bar g^{(3)} - 0.10\, 
\(D_{\mathbf{8}}^{(3)}-D_{\mathbf{1}}^{(3)}\)\)\,\textrm{pb},\\
\textrm{LHC$8$:}\; \sigma^{\mathrm{N^3LO}}_{\mathrm{approx}}\(\rho_h,m^2\)
&=\(253.98\pm 4.35 +0.14 \,\bar g^{(3)} -0.14\,
\(D_{\mathbf{8}}^{(3)}-D_{\mathbf{1}}^{(3)}\)\)\,\textrm{pb},\\
\textrm{LHC$13$:}\; \sigma^{\mathrm{N^3LO}}_{\mathrm{approx}}\(\rho_h,m^2\)
&=\(835.61\pm 14.78 +0.51\, \bar g^{(3)} -0.46\, 
\(D_{\mathbf{8}}^{(3)}-D_{\mathbf{1}}^{(3)}\)\)\,\textrm{pb},\\
\textrm{LHC$14$:}\; \sigma^{\mathrm{N^3LO}}_{\mathrm{approx}}\(\rho_h,m^2\)
&=\(988.57\pm 17.55 +0.61\, \bar g^{(3)} -0.54\, 
\(D_{\mathbf{8}}^{(3)}-D_{\mathbf{1}}^{(3)}\)\)\,\textrm{pb},
\end{align}
\end{subequations}
 where 
\beq
\label{eq:Hi}
\bar g^{(i)}=\bar g_{\mathbf{1}}^{(i)}+\bar g_{\mathbf{8}}^{(i)}.
\eeq
The constant $\bar g^{(3)}$ is not known.
The difference between
the coefficients $D_{\mathbf{8}}^{(3)}$ and $D_{\mathbf{1}}^{(3)}$ parametrizes the
missing information about the single log at N$^3$LO, as discussed in 
Section~\ref{sec:softres}.
Eq.~\eqref{eq:hqpapprox} is obtained using
exact expressions for the NLO~\cite{Czakon:2008ii} and 
NNLO~\cite{Czakon:2013goa,Baernreuther:2012ny,Czakon:2012pz,Czakon:2012zr},
combined with our approximation
for the N$^3$LO in the gluon channel.

\begin{table}[tb]
\centering
\begin{tabular} { c| c c c c}
& $D_{\mathbf{8}}^{(i)}-D_{\mathbf{1}}^{(i)}$ & $\bar g^{(i)}$ & $g^{(i)}$ & $g^{(i)}-\bar g^{(i)}$ \\
\midrule
$\as$   & $-0.955$ & $1.188$ & $3.832$ & $2.644$ \\
$\as^2$ & $-1.782$ & $0.535$ & $8.512$ & $7.977$ \\ 
$\as^3$ &     ?    &    ?    &    ?    & $12.348$ \\ 
\hline
\end{tabular}
\caption{The coefficients $D_{\mathbf{8}}^{(i)}-D_{\mathbf{1}}^{(i)}$,
  $\bar g^{(i)}$,  $g^{(i)}$ and the difference between the latter two
   at NLO and
  NNLO.}
\label{tab:compH3}
\end{table}

We now consider various sources of uncertainty. First, we consider the
uncertainty related to missing coefficients.
In Tab.~\ref{tab:compH3}, we list the values of the coefficients 
$\bar g^{(i)}$, and of
$D_{\mathbf{8}}^{(i)}-D_{\mathbf{1}}^{(i)}$ for the two known
orders, as well as the coefficients $g^{(i)}$ defined using the
analogue of Eq.~(\ref{eq:Hi}) but for the expansion coefficients of 
$g_\mathbf{I}(\as)$ Eq.~(\ref{eq:g0vsg0bar}) which is used  in the
standard resummed results of
Refs.~\cite{Bonciani:1998vc,Czakon:2013goa}, and the difference
between the two, which is known to a higher perturbative order. 
Based on these values, we note that the perturbative behaviour of 
the coefficients $\bar g^{(i)}$ appears to be rather more stable than that
of the coefficients  $g^{(i)}$, and we
reasonably expect the unknown 
coefficients, $\bar g^{(3)}$ and
$D_{\mathbf{8}}^{(3)}-D_{\mathbf{1}}^{(3)}$,
to be both of $\Ord(1)$. 
 We conservatively  estimate the uncertainty
related to these unknown coefficients as
\beq
\bar g^{(3)} = 0\pm 5,\qquad
D_{\mathbf{8}}^{(3)}-D_{\mathbf{1}}^{(3)} = 0\pm 10,
\eeq
which we add in quadrature to the approximation uncertainty of
Eq.~\eqref{eq:hqpapprox}.  We also note in passing that the $N$-soft approximation
assumes $g^{(3)}=0$, while our estimate corresponds to a value of
order ten for this constant, thereby partly explaining the difference
observed in Fig.~\ref{fig:Kfactor3} between our result and the $N$-soft
approximation, as discussed in the previous Section.

A second source of uncertainty is due to the fact that we only predict
the contribution of the gluon channel.
\begin{table}[tb]
\centering
\begin{tabular} {c | c c c}
LHC$8$ & $\sigma_{q\bar{q}} (pb)$ & $\sigma_{qg} (pb)$ & $\sigma_{gg} (pb)$ \\
\midrule
LO & $29.94$ & $0$ & $116.33$ \\
NLO & $4.17$ & $-0.45$ & $66.66$ \\
NNLO & $2.17$ & $-2.30$ & $26.36$ \\
\hline
\end{tabular}
\caption{Comparison of contributions to top pair production cross
  section from different channel at LHC 8 TeV at LO,
  NLO, NNLO.}
\label{tab:channels}
\end{table}
In Tab.~\ref{tab:channels}, we show the contributions to the 
top-pair production
cross section coming from different partonic channels at LO, NLO,
NNLO, for the specific case of LHC at $8$ TeV.  We note that gluon
fusion contribution is always dominant, and largely dominant at high
perturbative orders.  This remains true for other LHC energies (while
at the Tevatron the $q\bar q$ component becomes dominant). Therefore,
we expect that our estimate for the N$^3$LO contribution based on the
$gg$ channel only is only mildly affected by the lack of knowledge of
the other channels.
However, in order to take into account the uncertainty due to the missing
partonic channels, we consider the dependence of our result on the
factorization scale.  Indeed, all the scale dependent terms at N$^3$LO
are available (for all channels), but the inclusion of all of them
would consistently make the residual scale dependence of relative
order $\as^4$. On the other hand, if we include $\muf$ dependent terms
only in the $gg$ channel, the residual scale dependence is of order
$\as^3$, since it misses the compensation between channels. The
difference between these two ways of varying the scale can be thus
taken as an estimate of the size of the contribution from 
the missing quark channels.

\begin{figure}[t]
  \centering
  \includegraphics[width=0.495\textwidth]{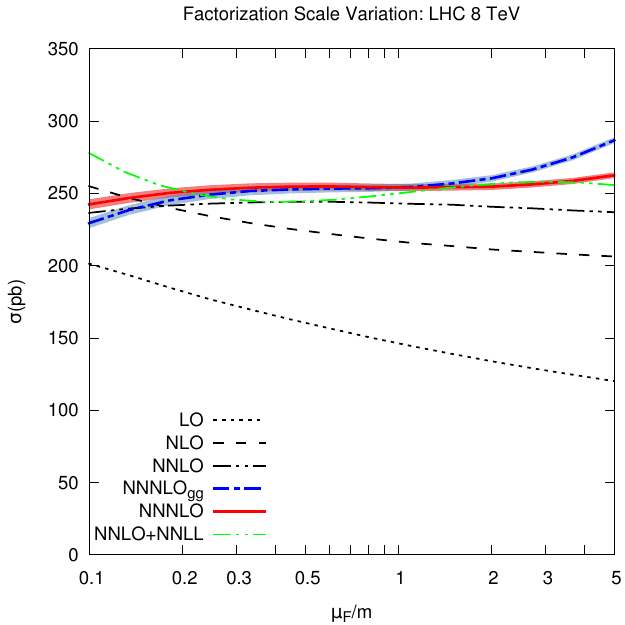}
  \includegraphics[width=0.495\textwidth]{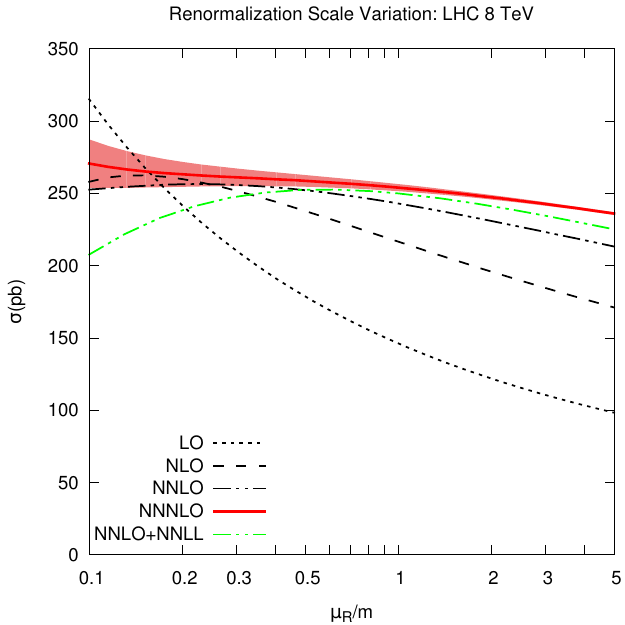}
  \caption{Factorization (left) and renormalization (right) scale
    dependence of the top production cross section at various
    perturbative orders. At N$^3$LO the factorization scale dependence
    is shown both including the contribution from all channels 
(NNNLO) and from the gluon channel only
    (NNNLO$_{gg}$). }
  \label{fig:scaledep}
\end{figure}
Finally, uncertainties related to missing higher-order terms can be
estimated by varying the renormalization and factorization scales in
the usual way.
The dependence of the cross section
at LO, NLO, NNLO and approximate N$^3$LO on the factorization 
and renormalization scales, is shown in a wide range in 
Fig.~\ref{fig:scaledep}; the NNLO+NNLL result of
Refs~\cite{Bonciani:1998vc,Czakon:2013goa} is also shown.  
The factorization scale variation 
at N$^3$LO with $\muf$ is shown both retaining contributions from all 
channels, and from the $gg$ channel only. 
The factorization scale dependence is rather mild
already at NNLO, and even milder at
approximate N$^3$LO when all channels are included.  When 
only the $gg$ channel is included, a stronger scale dependence is
observed, particularly at the extremes of the range, where sizable
contributions from the missing channels are generated.
The  renormalization scale dependence is somewhat stronger than the
factorization scale dependence, but  at N$^3$LO it has flattened out
almost completely, thereby indicating a good perturbative
convergence. It is interesting to observe that
the NNLO+NNLL result has a milder scale dependence than the NNLO, but
still a stronger scale dependence than the approximate  N$^3$LO. This
is what one may expect based on the observation that effectively,
because the process is far from threshold, the resummation is
providing some approximation mostly to N$^3$LO, which is however less
complete than the approximation which is constructed here.

Our final results, with  full uncertainty, are thus
\begin{subequations}
\label{eq:hqpapproxFinal}
\begin{align}
\textrm{LHC$7$:}\; \sigma^{\mathrm{N^3LO}}_{\mathrm{approx}}
&= 177.43 \,\textrm{pb} \pm 1.79\% \text{(approx)} \pm 0.97\% 
\text{(channels)} {}^{+3.02\%}_{-2.87\%} \text{(scales)}\\
\textrm{LHC$8$:}\; \sigma^{\mathrm{N^3LO}}_{\mathrm{approx}}
&= 253.98 \,\textrm{pb} \pm 1.82\% \text{(approx)} \pm 0.96\% \text{(channels)} {}^{+2.98\%}_{-2.83\%} \text{(scales)}\\
\textrm{LHC$13$:}\; \sigma^{\mathrm{N^3LO}}_{\mathrm{approx}}
&= 835.61 \,\textrm{pb} \pm 1.88\% \text{(approx)} \pm 0.96\% \text{(channels)} {}^{+2.73\%}_{-2.65\%} \text{(scales)}\\
\textrm{LHC$14$:}\; \sigma^{\mathrm{N^3LO}}_{\mathrm{approx}}
&= 988.57 \,\textrm{pb} \pm 1.88\% \text{(approx)} \pm 0.97\% \text{(channels)} {}^{+2.68\%}_{-2.62\%} \text{(scales)}
\end{align}
\end{subequations}
where the ``channels'' uncertainty has been computed as ($\pm$ half)
the difference between the $\muf$ scale variation evaluated with only
the $gg$ channel or with all the channels (NNNLO$_{gg}$ and NNNLO curves
in Fig.~\ref{fig:scaledep}), in the range $m/2\leq\muf\leq2m$ with
$\mur=m$.  The ``scales'' uncertainty is instead obtained through 
a canonical seven-point
variation, namely $m/2\leq \mur, \muf\leq 2m$ with
$1/2\leq\mur/\muf\leq2$, computed with all channels. We observe that
the approximation uncertainties, though conservatively estimated, 
are rather smaller than the scale
uncertainty, and in fact adding in  quadrature scale and
approximation uncertainties we end up with an overall theoretical
uncertainty on our N$^3$LO result of $3.5\%$, not much larger than the
scale uncertainty itself.
This uncertainties can be compared to the PDF uncertainty, which 
 at the LHC 
$\sqrt{s}=13$~TeV (with NNPDF 3.0 PDFs) is of order
$2\%$. Additional uncertainties come from the values of $\alpha_s$ and
$m_t$: see Ref.~\cite{Czakon:2013tha} for a more detailed discussion.

We observe that the uncertainty due to scale variations at NNLO
is about 5\% at the collider energies we are considering, which is
larger than the overall uncertainty on our
N$^3$LO estimate even accounting for approximation uncertainties.
The inclusion of our approximate N$^3$LO contribution
appears thus to be advantageous, and it leads to a decrease in
theoretical uncertainty which is of the size one would expect when
  going ftom NNLO to N$^3$LO.

We may  compare our approximate N$^3$LO result to that
which would be obtained using  the $N$-soft result shown 
in Fig.~\ref{fig:Kfactor3}. The latter leads to a   N$^3$LO
contribution which corresponds  to  an
increase of $3\%,2.8\%,2.3\%,2.4\%$ in comparison to the NNLO at LHC
$\sqrt{s}=7, 8
,13,14$~TeV, respectively. This is rather lower than our approximate
result, which corresponds to an increase of
$4.3\%,4.5\%,4.2\%,4.3\%$ respectively. The reasons for this have been
discussed in the previous section. As discussed above, and as it is
apparent from Fig.~\ref{fig:scaledep}, the scale uncertainty on this
result is somewhat smaller than that on the NNLO result, but larger
than that on our result.

Finally, we note that an approximate N$^3$LO result is also presentes in
Ref.~\cite{Kidonakis:2014isa}.  This result is obtained by considering
logarithmic terms enhanced at partonic threshold in the differential
cross section, inclusive in one of the two heavy quarks produced.  The
result is then integrated to obtain the inclusive cross section. The
enhancements over the NNLO results were found to be $4\%,3.6\%,2.7\%,2.6\%$
at $\sqrt{s}=7, 8 ,13,14$~TeV respectively. We leave a more detailed
comparison to the results of Ref.~\cite{Kidonakis:2014isa}, as well as
to the resummed result obtained in SCET
(e.g. Refs.~\cite{Beneke:2009ye,Beneke:2010da} and
Refs.~\cite{Ahrens:2010zv,Ahrens:2011px,Yang:2014hya}), for future
work.

\section{Conclusions}
\label{sec:conclusion}
We have constructed an approximate expression for the N$^3$LO
contribution to the production cross section of a heavy-quark pair at
hadron-hadron colliders. We have focused on the gluon-gluon initiated
subprocess, which gives the largest contribution at the LHC.

We have obtained our result by extending the method developed by some
of us for the case of Higgs production in gluon-gluon
fusion~\cite{Ball:2013bra}, and based on reconstructing the
Mellin-space partonic cross section from its known singularities.
Heavy-quark production is a process with a rather more complicated
kinematic and color structure compared to inclusive Higgs
production. Furthermore, fixed-order coefficient function for this
process are known to have a rather non-trivial singularity structure
in physical space, at first sight unrelated to the threshold or
high-energy limits. 
Therefore, application to this case of the the technique suggested 
in Ref.~\cite{Ball:2013bra} and applied there to Higgs production in
gluon fusion provides a rather stringent test of this methodology.
In this study we have shown  that the method
of Ref.~\cite{Ball:2013bra} provides excellent approximations to known
results up to NNLO, thereby validating he methodology.

Having established the reliability of the methodology even in this
more subtle case, we have used it to produce a N$^3$LO approximate
partonic cross section for heavy quark production.
We have then focused on the $t \bar t$ cross section, which
is of  great interest at the LHC. We have found that the
approximate N$^3$LO correction amounts to an increase in comparison to
the NNLO prediction of $4.3\%,4.5\%,4.2\%,4.3\%$, for $pp$ collisions at
$\sqrt{s}=7, 8 ,13,14$~TeV, respectively. Inclusion of this correction 
reduces the scale
uncertainty to 3\%, with a combined   uncertainty on the approximation itself  of
comparable size or smaller. Our final overall conservatively
estimated uncertainty is thus somewhat smaller than the uncertainty on
the exactly known NNLO result, and  inclusion of our approximate result
appears  to 
be advantageous.

\acknowledgments

We thank Alex Mitov for raising the issue discussed in
Sect.~\ref{sect:sings} and for several illuminating discussions.
SF and GR are supported in part by an Italian PRIN2010 grant, and SF
also by a European Investment Bank EIBURS grant, and by the European
Commission through the HiggsTools Initial Training Network
PITN-GA-2012-316704. SM is supported by the U.S.\ National Science
Foundation, under grant PHY--0969510, the LHC Theory Initiative.  MB
is supported by an European Research Council Starting Grant ``PDF4BSM:
Parton Distributions in the Higgs Boson Era''.

\appendix

\section{Coefficients in the large-$N$ contribution} \label{app:largeN}

The coefficients $b^{(A)}_{n,k}(\muf^2)$ and 
$b^{(D)}_{n,k,\mathbf{I}}(\muf^2)$ defined in Eq.~\eqref{eq:Sud} for $n\leq 3$ are given by
\begin{subequations}
\label{eq:bA}
\begin{align}
b^{(A)}_{1,0}&=4 A^{(1)}\ln 2 + 2 A^{(1)} \ln\frac{m^2}{\muf^2} \\
b^{(A)}_{1,1}&=4 A^{(1)}\\
b^{(A)}_{2,0}&=4 A^{(2)}\ln 2 - 4 A^{(1)} \beta_0 \ln^2 2 + 2 A^{(2)} \ln\frac{m^2}{\muf^2} + A^{(1)}\beta_0\ln^2\frac{m^2}{\muf^2}\\
b^{(A)}_{2,1}&=4 A^{(2)}-8 A^{(1)} \beta_0 \ln 2\\
b^{(A)}_{2,2}&=-4 A^{(1)} \beta_0\\
b^{(A)}_{3,0}&=4 A^{(3)}\ln 2-8 A^{(2)}\beta_0 \ln^2 2 -4 A^{(1)}\beta_1\ln^2 2+\frac{16}{3}A^{(1)}\beta^2_0\ln^3 2 \notag \\
&+2 A^{(3)}\ln\frac{m^2}{\muf^2}+\(2 A^{(2)}\beta_0+A^{(1)}\beta_1\)\ln^2\frac{m^2}{\muf^2}+\frac{2}{3}A^{(1)}\beta_0^2\ln^3\frac{m^2}{\muf^2}\\
b^{(A)}_{3,1}&=4 A^{(3)}-16 A^{(2)}\beta_0 \ln 2 -8 A^{(1)}\beta_1\ln 2 + 16 A^{(1)}\beta_0^2\ln^2 2\\
b^{(A)}_{3,2}&=-8 A^{(2)}\beta_0-4 A^{(1)}\beta_1+16 A^{(1)}\beta_0^2\ln 2\\
b^{(A)}_{3,3}&=\frac{16}{3} A^{(1)} \beta_0^2
\end{align}
\end{subequations}
and
\begin{subequations}
\label{eq:bD}
\begin{align}
b^{(D)}_{1,0,\mathbf{I}}&=D_{\mathbf{I}}^{(1)}\\
b^{(D)}_{2,0,\mathbf{I}}&=D_{\mathbf{I}}^{(2)}-2\beta_0 D_{\mathbf{I}}^{(1)}\ln 2\\
b^{(D)}_{2,1,\mathbf{I}}&=-2\beta_0D_{\mathbf{I}}^{(1)}\\
b^{(D)}_{3,0,\mathbf{I}}&=D_{\mathbf{I}}^{\(3\)}-2\beta_1 D_{\mathbf{I}}^{(1)} \ln 2 - 4 \beta_0 D_{\mathbf{I}}^{\(2\)}\ln 2 +4\beta_0^2 D_{\mathbf{I}}^{(1)}\ln^2 2\\
b^{(D)}_{3,1,\mathbf{I}}&=-2\beta_1 D_{\mathbf{I}}^{(1)}-4\beta_0 D_{\mathbf{I}}^{\(2\)}+8\beta_0^2 D_{\mathbf{I}}^{(1)}\ln 2\\
b^{(D)}_{3,2,\mathbf{I}}&=4\beta_0^2 D_{\mathbf{I}}^{(1)}
\end{align}
\end{subequations}
where
\begin{subequations}
\label{eq:A}
\begin{align}
A\(\as\)&=A^{(1)}\as+A^{(2)}\as^2+A^{(3)}\as^3+\Ord\(\as^4\);\\
\,&\,\notag\\
A^{(1)}&=\frac{\Ca}{\pi}\\
A^{(2)}&=\frac{\Ca}{2\pi^2}\(\(\frac{67}{18}-\zeta_2\)\Ca-\frac{5}{9}\Nf\)\\
A^{(3)}&=\frac{\Ca}{4\pi^3}\Big(\Ca^2\(\frac{245}{24}-\frac{67}{9}\zeta_2+\frac{11}{6}\zeta_3+\frac{11}{5}\zeta_2^2\)\notag\\
&+\Cf\Nf\(-\frac{55}{24}+2\zeta_3\)+\Ca\Nf\(-\frac{209}{108}+\frac{10}{9}\zeta_2-\frac{7}{3}\zeta_3\)-	\frac{\Nf^2}{27}\Big)
\end{align}
\end{subequations}
and
\begin{subequations}
\label{eq:D}
\begin{align}
D_{\mathbf{I}}\(\as\)&=D_{\mathbf{I}}^{(1)}\as+D_{\mathbf{I}}^{(2)}\as^2+D_{\mathbf{I}}^{(3)}\as^3+\Ord\(\as^4\);\\
\,&\,\notag \\
D_{\mathbf{1}}^{(1)}&=0\\
D_{\mathbf{1}}^{(2)}&=\frac{\Ca}{\pi^2}\(\(-\frac{101}{27}+\frac{11}{18}\pi^2+\frac{7}{2}\zeta_3\)\Ca+\(\frac{14}{27}-\frac{\pi^2}{9}\)\Nf\)\\
D_{\mathbf{1}}^{(3)}&=\frac{1}{\(4\pi\)^3}\Big(\Ca^3\(-\frac{594058}{729}+\frac{98224}{81}\zeta_2+\frac{40144}{27}\zeta_3-\frac{2992}{15}\zeta_2^2-\frac{352}{3}\zeta_2\zeta_3-384\zeta_5\)\notag\\
&+\Ca^2\Nf\(\frac{125252}{729}-\frac{29392}{81}\zeta_2-\frac{2480}{9}\zeta_3+\frac{736}{15}\zeta_2^2\)\notag\\
&+\Ca\Cf\Nf\(\frac{3422}{27}-32\zeta_2-\frac{608}{9}\zeta_3-\frac{64}{5}\zeta_2^2\)\notag\\
&+\Ca\Nf^2\(-\frac{3712}{729}+\frac{640}{27}\zeta_2+\frac{320}{27}\zeta_3\)\Big)\\
\,&\,\notag\\
D_{\mathbf{8}}^{(1)}&=D_{\mathbf{1}}^{(1)}-\frac{\Ca}{\pi}\\
D_{\mathbf{8}}^{(2)}&=D_{\mathbf{1}}^{(2)}+\frac{\Ca}{36\pi^2}\(\Ca\(-115+3\pi^2-18\zeta_3\)+22\Nf\),
\end{align}
\end{subequations}
while $D_{\mathbf{8}}^{(3)}$ is unknown.

To complete our resummed formula for the soft emission, we need
explicit expressions for the functions $\hat{\mathcal{D}}_k(N)$,
Mellin transforms of the distributions
\beq
\hat{\mathcal{D}}_k\(z\)=\plus{\frac{\ln^k\(1-z\)}{1-z}}+\left[\frac{\ln^k\frac{1-z}{\sqrt{z}}}{1-z}-\frac{\ln^k\(1-z\)}{1-z}\right],
\eeq
where the plus distribution is defined by
\beq
\int_0^1dz\,\plusq{f\(z\)}g\(z\)=\int_0^1dz\,g\(z\)\left[f\(z\)-f\(1\)\right].
\eeq
The full calculation is presented for example in
Ref.~\cite{Bonvini:2012sh, Ball:2013bra}. Here we give the final result:
\beq
\hat{\mathcal{D}}_k\(N\)=\frac{1}{k+1}\sum_{j=0}^k\binom{k+1}{j}\Gamma^{\(j\)}\(1\)\left[\Upsilon^{\(k+1-j\)}\(N,0\)-\Delta^{\(k+1-j\)}\(1\)\right],
\eeq
where we have defined
\begin{align}
\Delta\(\xi\)&=\frac{1}{\Gamma\(\xi\)}\\
\Upsilon\(N,\xi\)&=\Gamma\(N-\frac{\xi}{2}\)\Delta\(N+\frac{\xi}{2}\),
\end{align}
$\Gamma(N)$ is the usual Euler function, and the superscript in round brackets in $\Upsilon\(N,\xi\)$ denotes differentiation with respect to $\xi$.
For the first three values of $k$, relevant for our N$^3$LO approximation, we find
\begin{subequations}
\label{eq:Dkhat}
\begin{align}
\hat{\mathcal{D}}_0\(N\)&=-L,\\
\hat{\mathcal{D}}_1\(N\)&=+\frac12\[L^2+\zeta_2\],\\
\hat{\mathcal{D}}_2\(N\)&=-\frac13\[L^3+3\zeta_2L+2\zeta_3+\frac14\psi_2(N)\],\\
\hat{\mathcal{D}}_3\(N\)&=+\frac14\[L^4 + 6\zeta_2 L^2 + 8\zeta_3 L + \frac{27}5\zeta_2^2+\psi_2(N) L\],
\end{align}
\end{subequations}
with
\beq
L = \psi_0(N) + \gammae,
\eeq
where $\psi_k(N)=\psi_0^{(k)}(N)$, $\psi_0(N)=\Gamma'(N)/\Gamma(N)$ 
are the usual polygamma functions
and $\gammae=-\psi_0(1)$ is the Euler-Mascheroni constant.
 
The Coulomb functions $J_{\mathbf{I}}(N,\as)$ are computed by taking
a Mellin transform of the resummed momentum-space results obtained in the
context of pNRQCD~\cite{Beneke:2010da,Beneke:1999qg}. For more details
about the procedure of this Mellin transformation, see
Ref.~\cite{Muselli:tesi}.
The explicit expression for the Coulomb functions, up to N$^3$LO, 
is the following:
\begin{subequations}
\label{eq:J}
\begin{align}
J_{\mathbf{I}}(N,\as)&=1+J^{(1)}_{\mathbf{I}}(N)\as
+J^{(2)}_{\mathbf{I}}(N)\as^2+J^{(3)}_{\mathbf{I}}(N)\as^3+\Ord(\as^4)\\
&\notag \\
J_{\mathbf{1}}^{(1)}(N)&=\frac{\pi\Cf 2^{1-2N}}{B\(N,N\)}
\\
J_{\mathbf{1}}^{(2)}(N)&=\frac{\Cf^2\pi^2}{6}\(N-\frac{1}{2}\)
\notag\\
&-\Cf\(\Cf+\frac{\Ca}{2}\)\(-\gammae+2-2\(\psi_0\(2N\)-\psi_0\(N\)\)\)
\notag\\
&+\Bigg(\(-\frac{11}{24}\Cf\Ca+\frac{\Cf\Nf}{12}\)
\(\ln 2 +\frac{1}{2}\(-\gammae-\psi_0(N)\)\)
\notag\\
&+\(\frac{31}{72}\Cf\Ca-\frac{5}{36}\Cf\Nf\)\Bigg)
\frac{2^{2-2N}}{B\(N,N\)}
\\
J_{\mathbf{1}}^{(3)}(N)&=\frac{\Cf^2\pi}{108}\(N-\frac{1}{2}\)
\notag \\
&\(93-10\Nf-432\frac{\beta_0\zeta_3}{\pi}+36\pi\beta_0\(\gammae-2\ln 2+2\(\psi_0\(2N\)-\psi_0\(N\)\)\)\)
\notag\\
&-\frac{\Cf^2\pi}{108}+\Cf^2\(\Ca+2\Cf\)\frac{\pi}{2^{2N}B\(N,N\)}\(\gammae+\psi_0(N)\)\\
&\notag\\
J_{\mathbf{8}}^{(1)}(N)&=\frac{\pi\(\Cf-\frac{\Ca}{2}\)
2^{1-2N}}{B\(N,N\)}\\
J_{\mathbf{8}}^{(2)}(N)&=\frac{\(\Cf-\frac{\Ca}{2}\)^2\pi^2}{6}
\(N-\frac{1}{2}\)
\notag\\
&-\Cf\(\Cf-\frac{\Ca}{2}\)\(-\gammae +2-2\(\psi_0\(2N\)-\psi_0\(N\)\)\)\notag\\
&+\Bigg(\(-\frac{11}{24}\Ca+\frac{\Nf}{12}\)\(\Cf-\frac{\Ca}{2}\)
\(\ln 2 +\frac{1}{2}\(-\gammae-\psi_0(N)\)\)
\notag\\
&+\(\frac{31}{72}\Ca-\frac{5}{36}\Nf\)
\(\Cf-\frac{\Ca}{2}\)\Bigg)\frac{2^{2-2N}}
{B\(N,N\)}
\\
J_{\mathbf{8}}^{(3)}(N)&=\frac{\(\Cf-\frac{\Ca}{2}\)^2\pi}{108}\(N-\frac{1}{2}\)
\notag\\
&\(93-10\Nf-432\frac{\beta_0\zeta_3}{\pi}+36\pi\beta_0
\(\gammae-2\ln 2 +2\(\psi_0\(2N\)-\psi_0\(N\)\)\)\)
\notag\\
&-\frac{\(\Cf-\frac{\Ca}{2}\)^2\pi}{108}+\frac{\(2\Cf-\Ca\)^2\Cf\pi}{2^{2N-1}B\(N,N\)}
\(\gammae+\psi_0(N)\).
\end{align}
\end{subequations}
with $B\(a,b\)=\frac{\Gamma(a)\Gamma(b)}{\Gamma(a+b)}$ the Beta function.

We now turn to the functions $\bar g_{\mathbf{I}}(\as)$. They 
must be computed by matching with exact results, which are available up
to NNLO. Moreover, as pointed out in
Ref.~\cite{Czakon:2013goa}, at NNLO we can only infer from the exact result
the sum $\bar g^{(2)}=\bar g_{\mathbf{1}}^{(2)}+\bar g_{\mathbf{8}}^{(2)}$,
but not the individual contributions $\bar g^{(2)}_{\mathbf{I}}$.
The latter have been estimated by
the same procedure adopted in Ref.~\cite{Czakon:2013goa}, giving
\begin{subequations}
\label{eq:Hexplicit}
\begin{align}
&\bar g_{\mathbf{I}}(\as)=1+\bar g_{\mathbf{I}}^{(1)}\as+\bar g_{\mathbf{I}}^{(2)}
\as^2+\Ord\(\as^3\)\\
&\notag\\
&\bar g_{\mathbf{1}}^{(1)}=\frac{2}{7\pi}\[\Cf\(-5+\frac{\pi^2}{4}\)+
\Ca\(1+\frac{\pi^2}{12}\)\]\\
&\bar g_{\mathbf{1}}^{(2)}=0.216\\
&\notag\\
&\bar g_{\mathbf{8}}^{(1)}=\frac{5}{7\pi}\[\Cf\(-5+\frac{\pi^2}{4}\)+\Ca\(3
-\frac{\pi^2}{24}\)\]\\
&\bar g_{\mathbf{8}}^{(2)}=0.319.
\end{align}
\end{subequations}
It can be checked that the uncertainty due to this additional
guess is completely negligible in comparison to our error
bands. 
For simplicity we do not give the explicit factorization scale
dependence of $\bar g(\as)$.

\section{Coefficients in the small-$N$ contribution} 
\label{app:smallN}
The only ingredients which are needed for the
high-energy contribution are
the coefficients $h_{i_1,i_2}$ of the expansion of the impact factor,
up to N$^3$LO:
\begin{subequations}
\label{eq:himpact}
\begin{align}
h_{0,0}&=\frac{181\pi}{2160}\\
h_{1,0}&=h_{0,1}=\frac{7291\pi}{32400}\\
h_{1,1}&=\frac{502417\pi}{486000}-\frac{251\pi^3}{12960}\\
h_{2,0}&=h_{0,2}=\frac{58849\pi}{121500}\\
h_{2,1}&=h_{1,2}=\frac{\pi\(47041256-700575\pi^2+7526250\zeta_3\)}{14580000}\\
h_{3,0}&=h_{0,3}=\frac{\pi\(3608438+203625\zeta_3\)}{3645000}.
\end{align}
\end{subequations}
The coefficients $\bar{h}_{i_1,i_2}$ are simply obtained
dividing each $h_{i_1,i_2}$ by $h_{0,0}$. The factorization scale
dependence can be easily restored following the procedure explained for
example in Ref.~\cite{Ball:2013bra}.

The leading and next-to-leading singular contributions to the anomalous dimension
$\gamma^{\rm \sss +}$ are 
\begin{align}
\label{eq:gamma0}
\gamma^{(0)}&=\frac{e_{0,-1}}{N-1}+e_{0,0}+\Ord(N-1)\\
\label{eq:gamma1}
\gamma^{(1)}&=\frac{e_{1,-2}}{(N-1)^2}+\frac{e_{1,-1}}{N-1} + \Ord(1)\\
\label{eq:gamma2}
\gamma^{(2)}&=\frac{3_{2,-3}}{(N-1)^3}+\frac{e_{2,-2}}{(N-1)^2}
+\Ord\((N-1)^{-1}\),
\end{align}
with $e_{0,-1}=\frac{\Ca}{\pi}$, $e_{1,-2}=e_{2,-3}=0$ and 
\begin{align}
e_{0,0}&=\frac{-11\Ca^2+2\Nf(2\Cf-\Ca)}{12\pi\Ca}\\
e_{1,-1}&=\(\frac{13 \Cf}{18\pi^2}-\frac{23 \Ca}{36 \pi^2}\)\Nf\\
e_{2,-2}&=\frac{\Ca^3\zeta_3}{2\pi^3}+\frac{11\Ca^3\zeta_2}{12\pi^3}
-\frac{395\Ca^3}{108\pi^3}+\(\frac{\Ca^2\zeta_2}{6\pi^3}
-\frac{71\Ca^2}{108\pi^3}-\frac{\Cf\Ca\zeta_2}{3\pi^3}
+\frac{71\Cf\Ca}{54\pi^3}\)\Nf.
\end{align}

\end{document}